\documentclass[journal]{IEEEtran}
\usepackage{cite}
\usepackage{amsmath,amssymb,amsfonts}
\usepackage{algorithmic}
\usepackage{graphicx}
\usepackage{textcomp}
\def\BibTeX{{\rm B\kern-.05em{\sc i\kern-.025em b}\kern-.08em
    T\kern-.1667em\lower.7ex\hbox{E}\kern-.125emX}}
\usepackage{amssymb}
\usepackage{amsmath}
\usepackage{graphicx}
\usepackage{bbm}
\usepackage{booktabs}
\usepackage{multirow}
\usepackage{multicol}
\usepackage[flushleft]{threeparttable}
\usepackage{caption}
\usepackage{subfig}
\usepackage{stackengine}
\usepackage{makecell}
\usepackage{balance}
\usepackage{fancyhdr}

\usepackage[square,sort,comma,numbers]{natbib}

\begin{document}
\title{DeepMesh: Mesh-based Cardiac Motion Tracking using Deep Learning}


\author{Qingjie Meng,
        Wenjia Bai,
        Declan P O'Regan, 
        and Daniel Rueckert,~\IEEEmembership{Fellow, ~IEEE}
\thanks{This research has been conducted using the UK Biobank Resource under application number 40616. This work is supported by the British Heart Foundation (RG/19/6/34387, RE/18/4/34215); Medical Research Council (MC\_UP\_1605/13); National Institute for Health Research (NIHR) Imperial College Biomedical Research Centre. W. Bai is supported by EPSRC DeepGeM Grant (EP/W01842X/1); D. Rueckert is supported by ERC Advanced Grant Deep4MI (884622). (Declan O’Regan and Daniel Rueckert are joint senior authors). For the purpose of open access, the authors have applied a creative commons attribution (CC BY) licence to any author accepted manuscript version arising.}
\thanks{Q. Meng, W. Bai and D. Rueckert are with the Biomedical Image Analysis Group, Department of Computing, Imperial College London, SW7 2AZ, UK, (e-mail: q.meng16$\vert$w.bai$\vert$drueckert@imperial.ac.uk).}
\thanks{Corresponding author: Qingjie Meng}
\thanks{Q. Meng is also with School of Computer Science, University of Birmingham.}
\thanks{W. Bai is also with Department of Brain Sciences, Imperial College London.}
\thanks{D. P. O'Regan is with the MRC London Institute of Medical Sciences, Imperial College London, W12 0HS, UK. (e-mail: declan.oregan@imperial.ac.uk).}
\thanks{D. Rueckert is also with Klinikum rechts der Isar, Technical University Munich, Germany.}
}

\maketitle

\begin{abstract}
3D motion estimation from cine cardiac magnetic resonance (CMR) images is important for the assessment of cardiac function and the diagnosis of cardiovascular diseases. Current state-of-the art methods focus on estimating dense pixel-/voxel-wise motion fields in image space, which ignores the fact that motion estimation is only relevant and useful within the anatomical objects of interest, \emph{e.g.}, the heart. In this work, we model the heart as a 3D mesh consisting of epi- and endocardial surfaces. We propose a novel learning framework, \textit{DeepMesh}, which propagates a template heart mesh to a subject space and estimates the 3D motion of the heart mesh from CMR images for individual subjects. In \textit{DeepMesh}, the heart mesh of the end-diastolic frame of an individual subject is first reconstructed from the template mesh. Mesh-based 3D motion fields with respect to the end-diastolic frame are then estimated from 2D short- and long-axis CMR images. By developing a differentiable mesh-to-image rasterizer, \textit{DeepMesh} is able to leverage 2D shape information from multiple anatomical views for 3D mesh reconstruction and mesh motion estimation. 
The proposed method estimates vertex-wise displacement and thus maintains vertex correspondences between time frames, which is important for the quantitative assessment of cardiac function across different subjects and populations. We evaluate \textit{DeepMesh} on CMR images acquired from the UK Biobank.
We focus on 3D motion estimation of the left ventricle in this work. 
Experimental results show that the proposed method quantitatively and qualitatively outperforms other image-based and mesh-based cardiac motion tracking methods.
\end{abstract}


\section{Introduction}
\IEEEPARstart{E}{stimating} left ventricular (LV) myocardial motion is important for the detection of LV dysfunction and the diagnosis of myocardial diseases~\cite{Claus2015,Puyol2019}. 
Recent works utilize 3D surface meshes to represent the anatomy and assess the ventricular structure and function from meshes, \emph{e.g.}, quantifying pathological cardiac remodeling~\cite{Mansi2011} or characterizing LV motion phenotypes~\cite{Marvao2015}.
However, it remains a challenging problem to estimate cardiac motion on meshes directly from images, in particular, to keep the same mesh structure and vertex correspondence. 
Most recent cardiac motion tracking approaches utilize cine CMR images to estimate a dense motion field which represents pixel-/voxel-wise deformation in the image space, \emph{e.g.},~\cite{Qin2018,Bello2019,Puyol2019,Qin2020,Bai2020,Yu2020,Ye2021,Loecher2021,Meng2022_mulvimotion}.
Mapping the deformation from a pixel-/voxel-wise representation to a vertex-wise representation on a cardiac mesh is typically inefficient and can reduce the accuracy of motion estimation.
Specifically, a 2D pixel-wise motion field only considers the motion of the heart within a single view plane and does not provide complete 3D motion information. Using post-processing steps to convert 3D voxel-wise motion fields to 3D vertex-wise displacements may impair motion estimation accuracy due to interpolation. 

In this work, we propose a novel learning-based method \textit{DeepMesh} for estimating 3D cardiac motion on the heart mesh from 2D cine CMR images. The proposed method propagates a single template mesh to individual subjects and estimates both in-plane and through-plane motion on meshes by integrating information from short-axis (SAX) and long-axis (LAX) view images. Specifically, DeepMesh first utilizes a template heart mesh containing the epi- and endocardial surfaces to reconstruct the mesh at the end-diastolic (ED) frame for an individual heart from the input ED frame multi-view images. By deforming this template mesh, the proposed approach maintains the same mesh structure at the ED frame for all subjects. Subsequently, the multi-view images at the ED and $t$-th frames are utilized to directly estimate the 3D motion on the mesh. The estimated mesh motion explicitly shows the 3D displacement of each vertex from the ED frame to the $t$-th frame, and thus is able to maintain mesh structure and vertex correspondences between time frames. 
A differentiable mesh-to-image rasterizer is introduced during training to generate 2D soft segmentations from the 3D mesh. By comparing predicted 2D soft segmentations with ground truth 2D segmentations, the differentiable rasterizer allows leveraging of 2D multi-view anatomical shape information for both 3D mesh reconstruction and motion estimation. 
During inference, our model generates a sequence of meshes, which characterise the heart motion across the cardiac cycle.
Here, in this work, we model the left ventricle as a 3D mesh consisting of epi- and endocardial surfaces and estimate the LV myocardial motion.

\subsubsection*{Contributions}
This paper extends a preliminary version of the work presented at the MICCAI 2022 conference ~\cite{Meng2022_miccai}. In addition to the work in~\cite{Meng2022_miccai}, the main contributions in terms of methodology and evaluation are summarized as follows: 
\begin{itemize}

\item We additionally introduce a template-based mesh reconstruction module. This module reconstructs the ED frame mesh of individual subjects from a cardiac template and therefore, enables subsequent mesh-based motion tracking. With the mesh derived from the template, the proposed method is able to maintain the number of vertices and faces in the cohort. 

\item We add a new regularization term to the motion estimation module in~\cite{Meng2022_miccai}  and  demonstrate that this leads to an improved performance in motion tracking.

\item We conduct a more thorough experimental analysis of the proposed method. We quantitatively and qualitatively evaluate the performance of mesh reconstruction and mesh-based motion tracking. We additionally compare the proposed method with two state-of-the-art motion tracking methods which use multi-view images~\cite{Meng2022_mulvimotion, Meng2022_miccai}. Moreover, we perform an extensive ablation study with respect to anatomical views, loss combinations and hyper-parameter selections.

\end{itemize}


\section{Related work}

\subsection{Image-based motion estimation}

Many cardiac motion estimation methods, including conventional methods and deep learning-based methods, consider motion tracking within image space. They typically use image registration algorithms to estimate 2D pixel-wise or 3D voxel-wise motion fields.

\subsubsection{Conventional methods}
Image registration has been applied to cardiac motion estimation in previous works. For example, the free form deformation (FFD) method for non-rigid image registration~\cite{Rueckert1999} has been widely used for cardiac motion estimation in many recent works, \emph{e.g.},~\cite{Chandrashekara2003, Shen2005, Shi2012, Tobon2013, Puyol2018, Bello2019, Puyol2019, Bai2020}. De Craene et al.~\cite{Craene2012} introduced continuous spatio-temporal B-spline kernels for computing a 4D velocity field, which enforced temporal consistency in motion recovery.  Thirion~\cite{Thirion1998} developed the demons algorithm which utilizes diffusing models for image matching and further used it for cardiac motion tracking. Based on this work, Vercauteren et al.~\cite{Vercauteren2007} introduced a non-parametric diffeomorphic image registration method which has been used for cardiac motion tracking~\cite{Qin2020}.

\subsubsection{Deep learning-based methods}
In recent years, deep convolutional neural networks (CNNs) have inspired the exploration of deep learning-based cardiac motion estimation approaches~\cite{Duchateau2020}. 
Qin et al.~\cite{Qin2018} proposed a joint deep learning network for simultaneous cardiac segmentation and motion estimation. Their method contains a shared feature encoder which enables a weakly-supervised segmentation. The U-Net architecture~\cite{Ronneberger2015} has been widely used for learning-based image registration~\cite{Balakrishnan2019,XuZ2020} and further for cardiac motion estimation. For example, Zheng et al.~\cite{ZhengQ2019} proposed a method for cardiac pathology classification based on cardiac motion. This method utilizes a modified U-Net to generate flow maps between the ED frame and any other frame. Balakrishnan et al.~\cite{Balakrishnan2019} used 3D U-Net to build VoxelMorph for learning-based deformable image registration. Their registration method has been utilized in other cardiac motion tracking works, \emph{e.g.},~\cite{Ta2020}.  
Different from most of these previous deep learning-based methods that aim at 2D motion tracking by only using SAX stacks, Meng et al.~\cite{Meng2022_mulvimotion} focused on 3D motion tracking by fully combining multiple anatomical views. In their work, a deep learning model was proposed that learns 3D motion fields from a set of 2D SAX and LAX cine CMR images, which was able to estimate both in-plane and through-plane myocardial motion.
Regarding cardiac motion tracking in multiple datasets, Yu et al.~\cite{Yu2020} considered the distribution mismatch problem and proposed a meta-learning-based online model adaption framework. Towards motion tracking in tagged MRI images, Ye et al.~\cite{Ye2021} proposed a deep learning model where the motion fields between any two consecutive frames are first computed, and then combined to estimate the Lagrangian motion field between the ED frame and any other frame. Our method aims at 3D cardiac motion tracking from 2D images of multiple anatomical views. In contrast to~\cite{Meng2022_mulvimotion} which estimates 3D motion in image space, our method focuses on estimating 3D motion in mesh space. 

\begin{figure*}[pt]
 \centering
 \includegraphics[width=\textwidth, trim=3cm 9.3cm 8.5cm 0.3cm, clip]{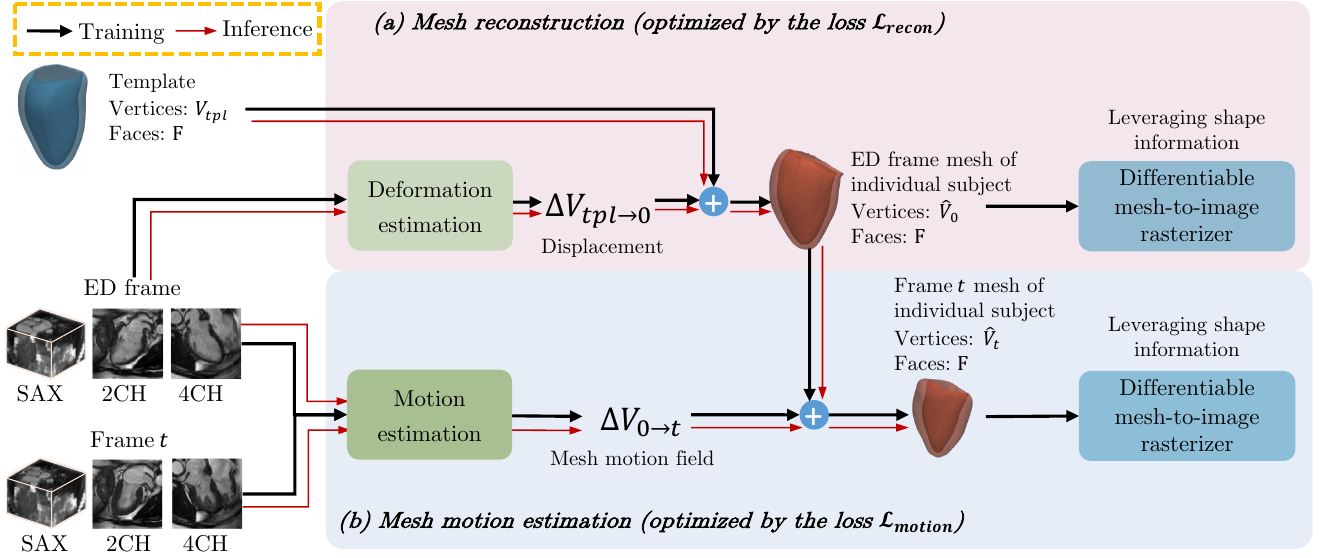}
 \caption{An overview of the proposed method. Panel (a) describes the mesh reconstruction module which reconstructs the ED frame mesh from a template mesh and the ED frame multi-view images. Panel (b) is the mesh motion estimation module, which takes multi-view images as input and learns 3D mesh motion field $\Delta V_{0\rightarrow t}$. By updating the reconstructed ED frame mesh with $\Delta V_{0\rightarrow t}$, the mesh of the $t$-th frame is predicted. During training, a differential mesh-to-image rasterizer is introduced to extract different 2D anatomical view planes from the predicted 3D meshes, which generates 2D soft segmentations. By comparing the predicted soft segmentations with ground truth segmentations, the rasterizer enables leveraging 2D shape information for 3D mesh reconstruction and motion estimation. Losses of each module are shown in Fig.~\ref{method_deform} and Fig.~\ref{method_motion}, accordingly.}
 \label{method_framework}
\end{figure*}

\subsection{Mesh-based motion estimation}

In contrast to dense motion estimation in image space, several other methods focus on anatomical motion estimation in mesh space~\cite{WangH2012}. These approaches explore mesh matching or mesh registration to estimate the motion field of the mesh. For example, Papademetris et al.~\cite{Papademetris2001} proposed a method that uses a biomechanical modeling and shape-tracking approach to estimate the motion of the myocardial mesh. Pan et al.~\cite{PanL2005} built a 3D mesh to represent material points inside the left ventricle wall and extended the 2D Harmonic phase (HARP) technique~\cite{Osman2000} to 3D for motion tracking of the mesh through a cardiac cycle. Abdelkhalek et al.~\cite{Abdelkhalek2020} built a framework to compute mesh displacements via point clouds alignment. These mesh motion estimation approaches compute mesh motion fields only from dynamic shape information, without considering intensity information from images. In contrast, our method combines image information with the myocardial mesh which contains the epi- and endocardial surfaces of the heart. We estimate 3D motion fields on meshes by using the intensity information of 2D images from multiple anatomical views.

\subsection{Mesh reconstruction}

In practice, the 3D mesh of the heart is not always available. Reconstructing a 3D mesh from images has been well investigated in the literature of general computer vision. Conventional approaches are based on multi-view geometry~\cite{MVG2003}. Although they can obtain high-quality reconstruction, these approaches are limited by the coverage provided by the multiple views. More recently, deep learning-based approaches are the major trend of 3D shape generation and they can reconstruct 3D meshes from only single or few images. Because of the difficulty of directly generating a feasible mesh structure, most learning-based methods learn shape priors from data and deform a sphere mesh to the target surface, \emph{e.g.},~\cite{Kanazawa2018,WangN2018,panJ2019,Gupta2020, shiY2021, Amiranashvili2022, Sun2022}.

In medical imaging, 3D shape reconstruction of the heart has been studied in the literature. For example, Villard et al.~\cite{Villard2018} proposed a data fitting method for cardiac surface reconstruction from 2D cardiac contours. This method iteratively optimizes the surface smoothness term and the contour matching term to obtain the 3D mesh of the heart. However, this method obtains meshes without maintaining vertex correspondences across the cohort.
Bello et al.~\cite{Bello2019} extracted heart surface meshes from image segmentations using the marching cube algorithm. This method also does not maintain vertex correspondences.
Romaszko et al.~\cite{Romaszko2021} proposed a deep neural network to predict point clouds of the heart from images. Following their work, Joyce et al.~\cite{Joyce2022} proposed a mesh fitting method which iteratively optimizes shape parameters (\emph{e.g.}, scalars, orientations) in order to match a mesh to the input 2D segmentations. 
Xia et al.~\cite{xia2022} proposed a method that uses CNNs for statistical shape modeling, in particular, adding phenotypic and demographic information for shape reconstruction. Their method estimates shape parameters and transformation parameters to deform the mean shape of the population for each subject. However, their method needs conventional registration algorithms to generate reference 3D shape information for model training, \emph{i.e.}, reference shape parameters and reference transformation parameters.
Different from these previous works, we build a deep neural network that directly predicts the 3D surface mesh of the heart at the ED frame by deforming a cardiac template according to the input 2D multi-view cine CMR images. Our method is able to reconstruct corresponding heart meshes across different subjects, i.e. with a consistent number of vertices and faces.  

\section{Method}

Give a set of CMR images, our goal is to propagate a single template mesh to all subjects and for individual subjects to estimate the heart motion on meshes across the cardiac cycle.
Our task is formulated as follows: Let $\{V_{tpl}, F\}$ denote the template mesh, $\{I_0^{sa}, I_0^{2ch}, I_0^{4ch}\}$ denote the 2D SAX, LAX 2-chamber (2CH) and LAX 4-chamber (4CH) view images of the heart at the ED frame and $\{I_t^{sa}, I_t^{2ch}, I_t^{4ch}\}$ denote the multi-view images at the $t$-th frame. $V_{tpl}$ and $F$ refer to the vertices and faces of the template mesh. $T$ is the number of frames in the cardiac cycle and $0\leqslant t\leqslant T-1$. We want to reconstruct the 3D heart mesh of individual subjects at the ED frame ($\{\hat{V}_0, F\}$) from the template, and then, for individual subjects, to estimate a 3D mesh motion field $\Delta V_{0\rightarrow t}$ between the ED and $t$-th frame by using the corresponding multi-view images. Here, $\Delta V_{0\rightarrow t}$ represents the motion of each vertex from the ED frame to the $t$-th frame, $\{V_{tpl},\hat{V}_0, \Delta V_{0\rightarrow t}\}\in\mathbb{R}^{N\times 3}$ and $N$ is the number of vertices.

\begin{figure*}[pt]
 \centering
 \includegraphics[width=\textwidth, trim=3cm 13cm 8.5cm 0.5cm, clip]{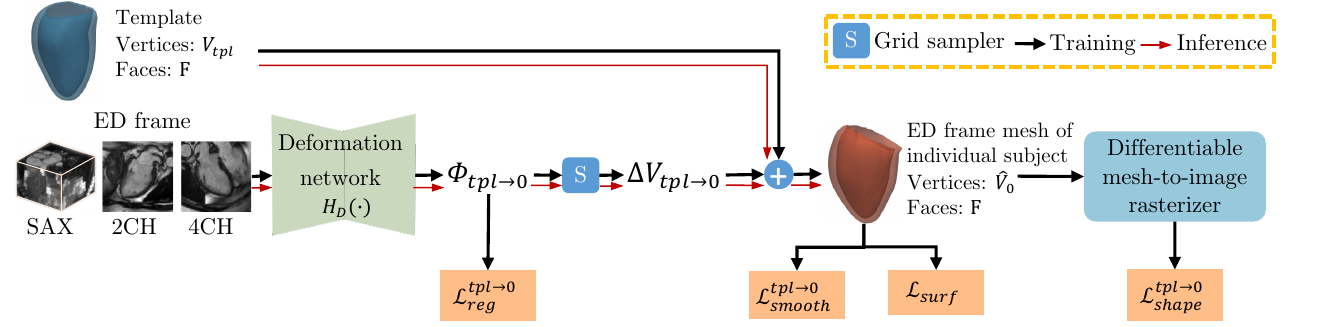}
 \caption{An overview of the mesh reconstruction module. This module reconstructs the ED frame mesh of individual subjects from a template mesh and multi-view images. In this module, the deformation network ($H_D(\cdot)$) predicts an intermediate voxel-wise displacement $\Phi_{tpl\rightarrow 0}$, and then $\Delta V_{tpl\rightarrow 0}$ containing the per-vertex displacement is generated by sampling from $\Phi_{tpl\rightarrow 0}$.}
 \label{method_deform}
\end{figure*}

The schematic architecture of the proposed method is shown in Fig.~\ref{method_framework}.
The proposed method can be separated into two main components: First, a mesh reconstruction module reconstructs the 3D mesh of the heart at the ED frame for individual subjects by deforming the template mesh (shown as the red box in Fig.~\ref{method_framework}). Second, a mesh motion estimation module learns the motion of a myocardial mesh from multi-view intensity images and deforms the ED frame mesh to the $t$-th frame based on the learned 3D mesh motion field (shown as the blue box in Fig.~\ref{method_framework}). During model training, a differentiable mesh-to-image rasterizer is introduced to yield 2D segmentations of the myocardium in the corresponding 2D planes (in the SAX and LAX orientations) by rasterizing the estimated 3D myocardial mesh. This enables using 2D segmentation information to supervise the mesh reconstruction and motion estimation modules.

\subsection{Mesh reconstruction}
\label{mesh_deformation}
This module aims to reconstruct the myocardial mesh for individual subjects at the ED frame. In particular, we leverage multi-view input images to learn a displacement $\Delta V_{tpl\rightarrow 0}$ that deforms the template mesh to the ED frame mesh of individual subjects vertex-by-vertex. Framework shown in Fig.~\ref{method_deform}.

\subsubsection{Deformation estimation}
\begin{figure}[t]
    \centering
    \subfloat[Deformation network $H_D(\cdot)$]{
    \includegraphics[height=6cm, trim=0cm 9cm 23cm 1cm, clip]{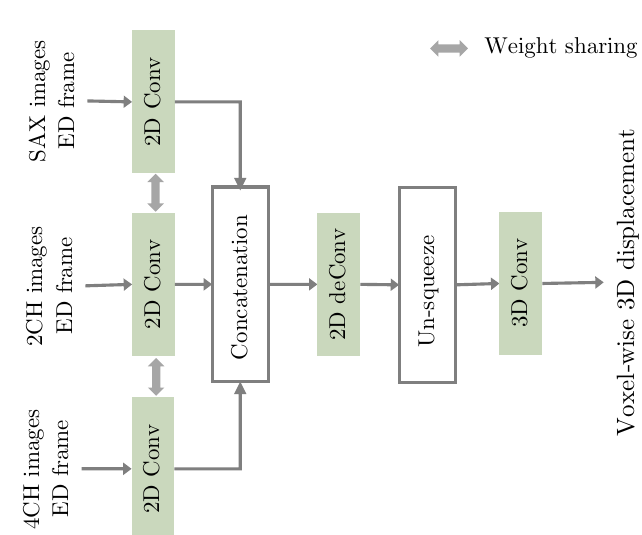} 
    }
    \\
    \subfloat[Motion network $H_M(\cdot)$]{
    \includegraphics[height=6.5cm, trim=0cm 8.5cm 18cm 0cm, clip]{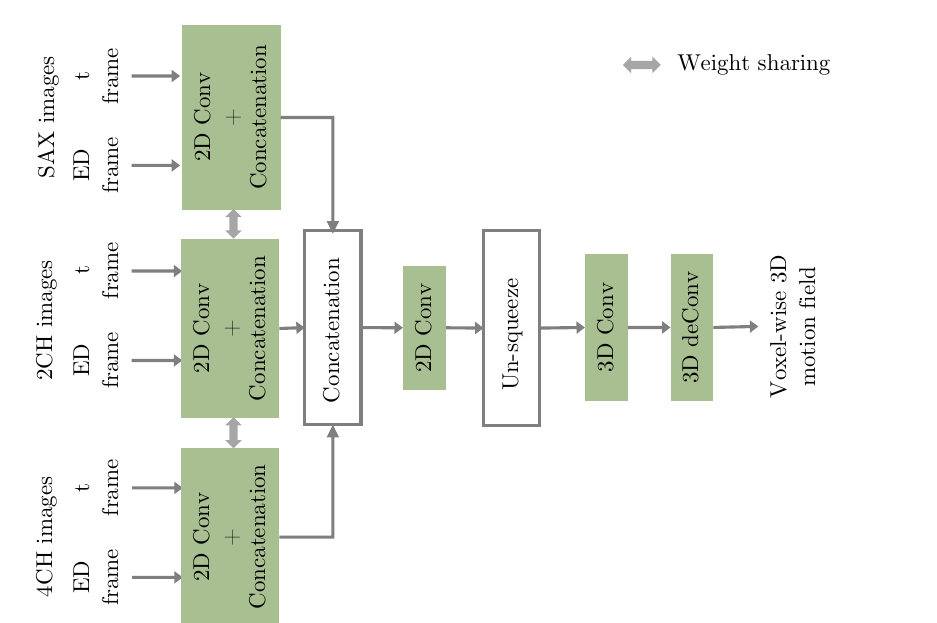} 
    } 
    \caption{A diagram of the network architecture of (a) the deformation network $H_D(\cdot)$ and (b) the motion network $H_M(\cdot)$. Here, \textit{Conv} represents convolutional layer with Relu and batch normalization while \textit{deConv} represents transposed convolutional layer with Relu and batch normalization. The detailed network architecture and code can be found in \textit{https://github.com/qmeng99/DeepMesh}.}
    \label{network_archi}
\end{figure}

We estimate $\Delta V_{tpl\rightarrow 0}$ from the input multi-view images of the ED frame. 
Specifically, a deformation network composed of a 2D CNN and a 3D CNN is introduced to learn an intermediate 3D voxel-wise displacement $\Phi_{tpl\rightarrow 0}$ from the 2D input SAX and LAX view images. The diagram of the deformation network architecture is shown in Fig.~\ref{network_archi} (a), where 2D convolutional layers learn 2D features from input images, followed by 3D convolutional layers that further learn 3D representations and predict $\Phi_{tpl\rightarrow 0}$.  
Subsequently, a grid sampler is utilized to generate $\Delta V_{tpl\rightarrow 0}$ from the obtained $\Phi_{tpl\rightarrow 0}$. In detail, for each vertex of the input template, its displacement is sampled from $\Phi_{tpl\rightarrow 0}$ by using bi-linear interpolation at the coordinates of this vertex. Therefore, $\Delta V_{tpl\rightarrow 0}$ contains the displacement of each vertex from the template mesh to the ED frame mesh. 

We formulate the deformation estimation as follows,
\begin{equation}\label{estimateV0}
\Delta V_{tpl\rightarrow 0}=S(H_D(I_0^{sa}, I_0^{2ch}, I_0^{4ch}), V_{tpl}).
\end{equation}
Here, $H_D(\cdot)$ is the deformation network, $S(\cdot, \cdot)$ is the grid sampler and $\Phi_{tpl\rightarrow 0}=H_D(I_0^{sa}, I_0^{2ch}, I_0^{4ch})$.

\subsubsection{Reconstructing the ED frame mesh}
With the estimated $\Delta V_{tpl\rightarrow 0}$, the ED frame mesh ($\{\hat{V}_0, F\}$) of individual subject can be reconstructed by deforming the input template ($\{V_{tpl}, F\}$),
\begin{equation}\label{mesh_recon}
\hat{V}_0=V_{tpl}+\Delta V_{tpl\rightarrow 0}.
\end{equation}

A Laplacian smoothing loss\footnote{Implemented by pytorch3d.loss.mesh$\_$laplacian$\_$smoothing()\label{laplaciansmooth}} $\mathcal{L}^{tpl\rightarrow 0}_{smooth}$ is used to evaluate the smoothness of the reconstructed ED frame mesh. The Laplacian of a vertex $\hat{v}_0^i$ is defined by $L(\hat{v}_0^i)$,
\begin{equation}\label{loss_smooth_v0}
L(\hat{v}_0^i)=\frac{1}{|\mathcal{N}_i|}\sum_{j\in \mathcal{N}_i}(\hat{v}_0^i-\hat{v}_0^j).
\end{equation}
Here, $\{\hat{v}_0^i,\hat{v}_0^j\}$ are vertices on $\hat{V}_0$ and $\mathcal{N}_i$ is the set of adjacent vertices to $\hat{v}_0^i$.

A surface loss $\mathcal{L}_{surf}$ penalizes the similarity between the reconstructed mesh ($\{\hat{V}_0, F\}$) and the ground truth mesh ($\{V_0, F\}$) of the ED frame. We use the Chamfer distance\footnote{Implemented by pytorch3d.loss.chamfer$\_$distance()} as the implementation,
\begin{equation}\label{loss_cd}
\scalebox{0.9}{$\mathcal{L}_{surf}=\frac{1}{|\hat{V}_0|}\sum\limits_{\hat{v}^i_0\in \hat{V}_0}\min\limits_{v^j_0\in V_0}\|\hat{v}_0^i-v_0^j\|^2_2+\frac{1}{|V_0|}\sum\limits_{v^j_0\in V_0}\min\limits_{\hat{v}^i_0\in \hat{V}_0}\|\hat{v}_0^i-v_0^j\|^2_2$}.
\end{equation}

In addition, we utilize the Huber loss used in~\cite{Qin2018,Meng2022_mulvimotion} as a regularization term to encourage a smooth intermediate $\Phi_{tpl}$,
\begin{equation}\label{loss_reg_V0}
\mathcal{L}^{tpl\rightarrow 0}_{reg}= \sqrt{\epsilon+\sum_{i=1}^{Q}\|\triangledown\Phi_{tpl}(q_i)\|^2}.
\end{equation}
Same to~\cite{Qin2018, Meng2022_mulvimotion}, $\epsilon$ is set to 0.01. $q_i$ is the $i$-th voxel and $Q$ denotes the number of voxels. 

As we aim to learn a 3D dense deformation from 2D sparse images for mesh reconstruction, the current losses have difficulty to guarantee accurate performance. To address this problem, we introduce a shape constraint from 2D segmentations as an additional regularization. This regularization term $\mathcal{L}^{tpl\rightarrow 0}_{shape}$ is described in detail in Sec.~\ref{diffmesh2img}.

\subsection{Mesh motion estimation}
\label{meshmotion_estimation}
\begin{figure*}[pt]
 \centering
 \includegraphics[width=\textwidth, trim=2.5cm 11.5cm 8.5cm 0.7cm, clip]{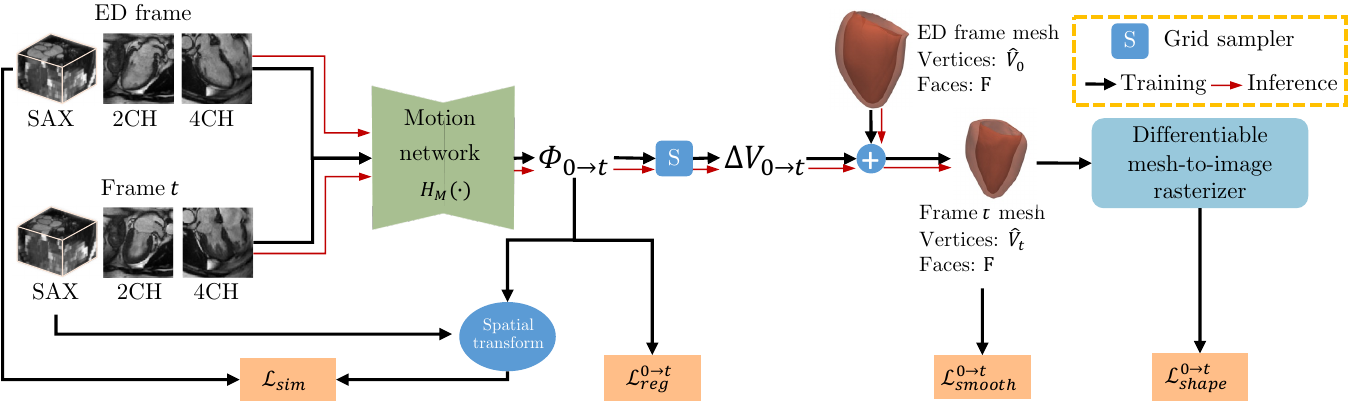}
 \caption{An overview of the mesh motion estimation module. This module estimates the motion of the heart mesh from the ED frame to the $t$-th frame. It takes multi-view images of the ED frame and the $t$-th frame as input and learns vertex-wise 3D mesh motion field $\Delta V_{0\rightarrow t}$ via predicting an intermediate voxel-wise motion field $\Phi_{0\rightarrow t}$. By updating the myocardial mesh of the ED frame with $\Delta V_{0\rightarrow t}$, the mesh of the $t$-th frame is predicted.}
 \label{method_motion}
\end{figure*}

In this module, we take multi-view images of the ED frame and the $t$-th frame as input to estimate a vertex-wise 3D mesh motion field $\Delta V_{0\rightarrow t}$. Then, we predict the mesh at the $t$-th frame by deforming the ED frame mesh reconstructed in the previous module using the 3D motion field $\Delta V_{0\rightarrow t}$. Fig.~\ref{method_motion} shows the overview of this module. 

\subsubsection{Motion estimation}
We estimate $\Delta V_{0\rightarrow t}$ from the input images via predicting an intermediate voxel-wise 3D motion field $\Phi_{0\rightarrow t}$. 
In detail, we build a motion network which consists of a 2D CNN and a 3D CNN to first learn $\Phi_{0\rightarrow t}$.
This motion network combines 2D multi-view images at both the ED frame and the $t$-th frame to estimate the intermediate 3D voxel-wise motion field $\Phi_{0\rightarrow t}$. The diagram of the motion network architecture is in Fig.~\ref{network_archi} (b), where 2D convolutional layers learn 2D features from two time frames and 3D convolutional layers predict $\Phi_{0\rightarrow t}$. 
The obtained $\Phi_{0\rightarrow t}$ represents the motion of image voxels from the ED frame to the $t$-th frame.
Then, a grid sampler is utilized to generate $\Delta V_{0\rightarrow t}$ from the obtained $\Phi_{0\rightarrow t}$ based on the vertices on the reconstructed ED frame mesh ($\hat{V}_0$) and bi-linear interpolation. $\Delta V_{0\rightarrow t}$ represents the motion field of each vertex from the ED frame to the $t$-th frame. Overall, $\Delta V_{0\rightarrow t}$ is estimated from the input multi-view images by
\begin{equation}\label{estVt}
\Delta V_{0\rightarrow t}=S(H_M(I_0^{sa}, I_0^{2ch}, I_0^{4ch}, I_t^{sa}, I_t^{2ch}, I_t^{4ch}), \hat{V}_0).
\end{equation}
Here, $H_M(\cdot, \cdot)$ is the motion network and $\Phi_{0\rightarrow t}=H_M(\cdot)$. 

\subsubsection{Mesh prediction}
With the estimated $\Delta V_{0\rightarrow t}$, the reconstructed ED frame mesh ($\{\hat{V}_0, F\}$) can be deformed to the $t$-th frame ($\{\hat{V}_t, F\}$) by
\begin{equation}\label{mesh_pred}
\hat{V}_t=\hat{V}_0+\Delta V_{0\rightarrow t}.
\end{equation}

As ground truth mesh displacement is usually unavailable, $\Delta V_{0\rightarrow t}$ can not be directly evaluated. Instead, we evaluate $\Phi_{0\rightarrow t}$ in a self-supervised manner. We transform the SAX stack of the $t$-th frame ($I_t^{sa}$) to the ED frame using $\Phi_{0\rightarrow t}$ via a spatial transformer network~\cite{Jaderberg2015}. By minimizing the image similarity loss in Eq.~\ref{loss_sim}, $\Phi_{0\rightarrow t}$ is encouraged to reflect the motion of the myocardium. 
\begin{equation}\label{loss_sim}
\mathcal{L}_{sim}=\|I_0^{sa}-I_t^{sa}\circ\Phi_{0\rightarrow t}\|^2
\end{equation}

Similar to Eq.~\ref{loss_smooth_v0}, the smoothness of the predicted $t$-th frame mesh is evaluated by a Laplacian smoothing loss$^{\ref{laplaciansmooth}}$ $\mathcal{L}^{0\rightarrow t}_{smooth}$. 

The gradients of the intermediate $\Phi_{0\rightarrow t}$ are penalized by the Huber loss similar to Eq.~\ref{loss_reg_V0}, $\mathcal{L}^{0\rightarrow t}_{reg}=\sqrt{\epsilon+\sum_{i=1}^{Q}\|\triangledown\Phi_{0\rightarrow t}(q_i)\|^2}$.

For mesh motion estimation, we also introduce a shape constraint to better learn 3D dense deformation from 2D sparse images. This regularization term ($\mathcal{L}^{0\rightarrow t}_{shape}$) is described in detail in Sec.~\ref{diffmesh2img}.

\subsection{Differentiable mesh-to-image rasterizer}
\label{diffmesh2img}


As ground truth 3D deformation is usually unavailable, we want to use 2D anatomical shape information to further supervise both 3D mesh reconstruction and motion estimation. To achieve this, we propose a differentiable mesh-to-image rasterizer to extract 2D soft contours of the myocardium from the predicted 3D heart mesh at the ED frame and the $t$-th frame. By comparing with the ground truth 2D myocardial contours, the differentiable rasterizer enables using sparse 2D shape information from multiple views to supervise 3D mesh reconstruction and motion estimation.

\begin{figure}[pt]
 \centering
 \includegraphics[width=0.48\textwidth, trim=3.9cm 13.7cm 16cm 0.8cm, clip]{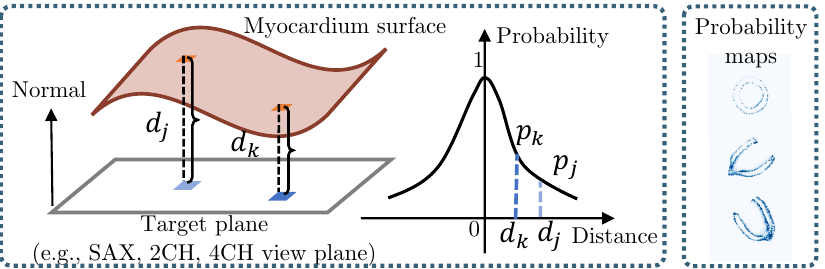}
 \caption{Illustration of the differentiable mesh-to-image rasterizer. This rasterizer extracts an anatomical view plane from a 3D mesh, and thus generates a 2D soft segmentation. In the left panel, $d_j$ and $d_k$ show the distance between a vertex of the heart surface and the target anatomical view plane. $p_j$ and $p_k$ refer to the probability of the vertex lying on the plane. The higher the distance, the lower the possibility the vertex lying on the plane. The right panel show examples of the obtained probability maps (2D soft contours).}
 \label{method_dmir}
\end{figure}


The input of the differentiable rasterizer is the predicted 3D mesh of the myocardium $\{\hat{V}_s, F\}$. The outputs are 2D contours of the myocardium intersecting on the SAX, 2CH and 4CH view planes ($\{P_s^{sa}, P_s^{2ch}, P_s^{4ch}\}$). Here $s=\{0,t\}$ refers to the ED frame and the $t$-th frame, respectively. When extracting a 2D plane from the 3D mesh, the vertices on the 3D mesh may not perfectly lie in the 2D plane. Therefore, we compute the probability of vertices lying on the plane, which is important for maintaining the differentiability. Specifically, we use probability maps to represent the 2D soft contours of the myocardium. Each pixel on the probability map represents the probability of a vertex from the 3D myocardial mesh lying on a specific 2D plane. The closer a vertex to a plane, the higher probability the vertex lies on the plane. Fig.5 illustrates the rasterizer. 

In detail, the coordinates of a vertex $\hat{v}_s^i$ ($\hat{v}_s^i\in\hat{V}_s, i=[0,1,...,N]$) are first transformed to the image space of different anatomical planes using the information about the relative position in the DICOM header of 2D images, \emph{e.g.}, $(x^{ik}_s, y^{ik}_s, z^{ik}_s)$ is the transformed coordinates of $\hat{v}_s^i$ and $k$ is the target 2D plane. Then, the probability of each vertex being on plane $k$ is estimated according to their distance:
\begin{equation}\label{soft_slicer}
p_s^{ik}=e^{-\tau {(d^{ik}_s)^2}}, \quad d^{ik}_s=|z^{ik}_s-z^k|, \quad k=\{{{sa}}, {2ch}, {4ch}\}
\end{equation}
Here $p_s^{ik}$ refers to the probability of $\hat{v}_s^i$ belonging to the plane $k$ and $\tau$ is the hyper-parameter which controls the sharpness of the exponential function. $d_s^{ik}$ is the distance between $\hat{v}_s^i$ and the plane $k$, and $z^k$ is the slice corresponding to the plane $k$. The vertices satisfying $d_s^{ik}<1$ are selected as the intersection of 3D mesh $\{\hat{V}_s, F\}$ and 2D plane $k$. The probability values of these vertices form the probability map $P_s^{k}$.


The obtained 2D probability maps are compared to 2D ground truth binary segmentations $\{B_s^{sa}, B_s^{2ch}, B_s^{4ch}\}$. Here, only ground truth contours of the myocardium are used and we compare between contours.
We utilize a weighted Hausdorff distance\footnote{Implemented by https://github.com/javiribera/locating-objects-without-bboxes}($\textrm{WHD}(\cdot,\cdot)$)~\cite{Ribera2019} to measure the similarity between these contours. $\mathcal{L}^{tpl\rightarrow 0}_{shape}$ is the shape regularization term for the mesh reconstruction module,
\begin{equation}\label{loss_whd}
\mathcal{L}^{tpl\rightarrow 0}_{shape}=\sum\nolimits_{k=\{sa,2ch,4ch\}}\textrm{{WHD}}(P_0^k, B_0^k).
\end{equation}

$\mathcal{L}^{0\rightarrow t}_{shape}$ is the shape regularization term for the mesh motion tracking module. It is the same format to Eq.~\ref{loss_whd} but the input are $P_t^k, B_t^k$.

As we use an exponential function (Eq.~\ref{soft_slicer}) for the rasterization, when minimizing the loss function (\emph{e.g.}, Eq.~\ref{loss_whd}), the gradient can be back-propagated to train the networks. Therefore, the exponential function enables the differentiability of the rasterization, and thus enables end-to-end model training.


\subsection{Optimization}
Our model is trained by two stages. The first stage is to train the mesh reconstruction module (\emph{i.e.}, Deformation Network $H_D(\cdot)$) by minimizing $\mathcal{L}_{recon}$ (Eq.~\ref{Loss_deform}). The inputs are the template mesh and the multi-view images at the ED frame. The output is the vertex-wise displacement which deforms the template mesh to the individual subject.
\begin{equation}\label{Loss_deform}
\mathcal{L}_{recon} = \mathcal{L}^{tpl\rightarrow 0}_{shape}+\lambda_1\mathcal{L}^{tpl\rightarrow 0}_{smooth}+\beta_1\mathcal{L}_{surf}+\gamma_1\mathcal{L}^{tpl\rightarrow 0}_{reg}.
\end{equation}
The second stage is to train the mesh motion estimation module (\emph{i.e.}, Motion Network $H_M(\cdot)$) by minimizing $\mathcal{L}_{motion}$ (Eq.~\ref{Loss_motion}). The inputs are the multi-view images of the ED frame and frame t. The output is the mesh motion field. For each training iteration, frame t is randomly selected from the cardiac cycle.
\begin{equation}\label{Loss_motion}
\mathcal{L}_{motion} = \mathcal{L}^{0\rightarrow t}_{shape}+\lambda_2\mathcal{L}^{0\rightarrow t}_{smooth}+\beta_2\mathcal{L}_{sim}+\gamma_2\mathcal{L}^{0\rightarrow t}_{reg}. 
\end{equation}
Here, $\{\lambda_i, \beta_i, \gamma_i\}_{i=\{1,2\}}$ are hyper-parameters chosen experimentally depending on the dataset. We use the Adam optimizer ($\text{learning rate}=10^{-4}$) to update the network parameters. Our model is implemented by Pytorch and is trained on a NVIDIA RTX A5000 GPU with 24GB of memory.

\section{Experiments}
We evaluate the performance of 3D mesh reconstruction and mesh motion tracking on the LV myocardium. We compare the proposed method, named as DeepMesh, with other image-based and mesh-based cardiac motion tracking methods. We explore the effectiveness of different loss components and the influence of the hyper-parameters. We show the key results in the main paper\footnote{Code is at DOI: 10.5281/zenodo.8200635}. The dynamic motion tracking videos can be found in \textit{https://github.com/qmeng99/DeepMesh}. 

\subsection{Experiment setups}
\subsubsection{Data} Experiments were performed on randomly selected 530 subjects from the UK Biobank study~\cite{Petersen2015}. Each subject contains SAX, 2CH and 4CH view cine CMR sequences and each sequence contains 50 frames. SAX view images were resampled by linear interpolation from a spacing of $\sim 1.8\times 1.8\times 10mm$ to a spacing of $1.25\times 1.25\times 2mm$ while 2CH and 4CH view images were resampled from $\sim 1.8\times 1.8 mm$ to $1.25\times 1.25mm$. Based on the center of the intersecting line between the middle slice of the SAX stack and the LAX view images, the SAX, 2CH and 4CH view images are cropped to cover the whole LV in the center. The input LV template mesh is provided by~\cite{Bai2015}. This template contains $22,043$ vertices and $43,840$ faces. For model training, 2D segmentations are used to supervise mesh reconstruction and motion tracking. The 2D binary segmentations used in Eq.~\ref{loss_whd} were extracted from a 3D high resolution segmentation. This 3D high resolution segmentation is generated via an automated tool provided in~\cite{Duan2019}, followed by manual quality control. We use 3D myocardial meshes of the ED frame and the end-systolic (ES) frame for evaluation. These ground truth 3D meshes are reconstructed from the 3D high resolution segmentations using the marching cube algorithm. We split the dataset into 400/50/80 for train/validation/test and train the proposed model for 300 epochs. 
We choose the hyper-parameters using grid search and select the hyper-parameters with the best performance on the validation data. Specifically, the hyper-parameters in Eq.~\ref{Loss_deform} are chosen from $\lambda_1=[10,20,30,40,50]$, $\beta_1=[0.1,0.3,0.5,0.7,0.9]$ and $\gamma_1=[0.1,0.3,0.5,0.7,0.9]$, and are selected as $\lambda_1=20, \beta_1=0.5, \gamma_1=0.5$. In Eq.~\ref{Loss_motion}, the hyper-parameters are chosen from $\lambda_2=[100,130,150,170,190]$, $\beta_2=[10,20,30,40,50]$ and $\gamma_2=[0.1,0.3,0.5,0.7,0.9]$ and are selected as $\lambda_2=150, \beta_2=20, \gamma_2=0.5$. In Eq.~\ref{soft_slicer}, we select $\tau=3$ from $\tau=[2,3]$.

\subsubsection{Evaluation metrics}
For evaluating the performance of 3D motion tracking on meshes, we compared the predicted 3D mesh and the ground truth 3D mesh at the ES frame. In addition, we extract 2D contours of the myocardium at SAX and LAX view planes from the predicted 3D meshes, and then compare the extracted 2D contours with the ground truth 2D contours (extracted from ground truth 3D meshes). 
The following metrics are used for evaluation: Surface distance, Hausdorff distance (HD) and Boundary F-score (BoundF). 
The surface distance evaluates the distance between the predicted and the ground truth meshes. The Hausdorff distance and Boundary F-score compare the predicted and the ground truth 2D myocardium contours on SAX, 2CH and 4CH view planes. The Hausdorff distance quantifies the contour distance while Boundary F-score evaluates contour alignment accuracy as described in~\cite{Perazzi2016,Cheng2019,Gur2020}. Here, to compute the Hausdorff distance at the SAX view, we average the Hausdorff distance of the second slice (slice 1), the middle slice (slice 4) and the second last slice (slice 7).

\subsubsection{Baseline methods}
We compared the proposed method with five state-of-the-art cardiac motion tracking approaches, including two conventional methods and three learning-based methods. The two conventional methods are a B-spline free form deformation (FFD) algorithm\footnote{Implemented by using the MIRTK toolkit: http://mirtk.github.io/}~\cite{Rueckert1999} and a diffeomorphic Demons (dDemons) algorithm\footnote{ https://github.com/InsightSoftwareConsortium/SimpleITK-Notebooks/Python/66\_Registration\_Demons.ipynb}~\cite{Vercauteren2007} which have been used in many recent cardiac motion tracking works~\cite{Tobon2013, Puyol2018, Bello2019, Puyol2019, Bai2020, Qin2020}. For the learning-based method, the UNet architecture has been used in many recent works for image registration~\cite{Balakrishnan2019, XuZ2020, Ta2020}, and thus our third baseline is a deep learning method with 3D-UNet\footnote{https://github.com/wolny/pytorch-3dunet}~\cite{ociek2016}. In addition, we compared the proposed method with MulViMotion\footnote{https://github.com/qmeng99/Multiview-Motion-Estimation-for-3D-cardiac-motion-tracking}~\cite{Meng2022_mulvimotion} and MeshMotion~\cite{Meng2022_miccai} which are two deep learning-based methods that utilize multi-view cardiac CMR images for 3D motion tracking. For fair comparison, we evaluated several sets of hyper-parameter values for all methods and selected hyper-parameters that achieve the best Hausdorff distance on the validation set. 

\subsection{Mesh-based motion tracking}

\subsubsection{Mesh reconstruction performance}
\begin{figure}[t]
    \centering
    \setcounter{subfigure}{0}
    \hspace{-1cm}
    \subfloat[Meshes]{
    \begin{tabular}{c}
         \includegraphics[height=1.9cm, trim=13.5cm 1.5cm 14cm 3.5cm, clip]{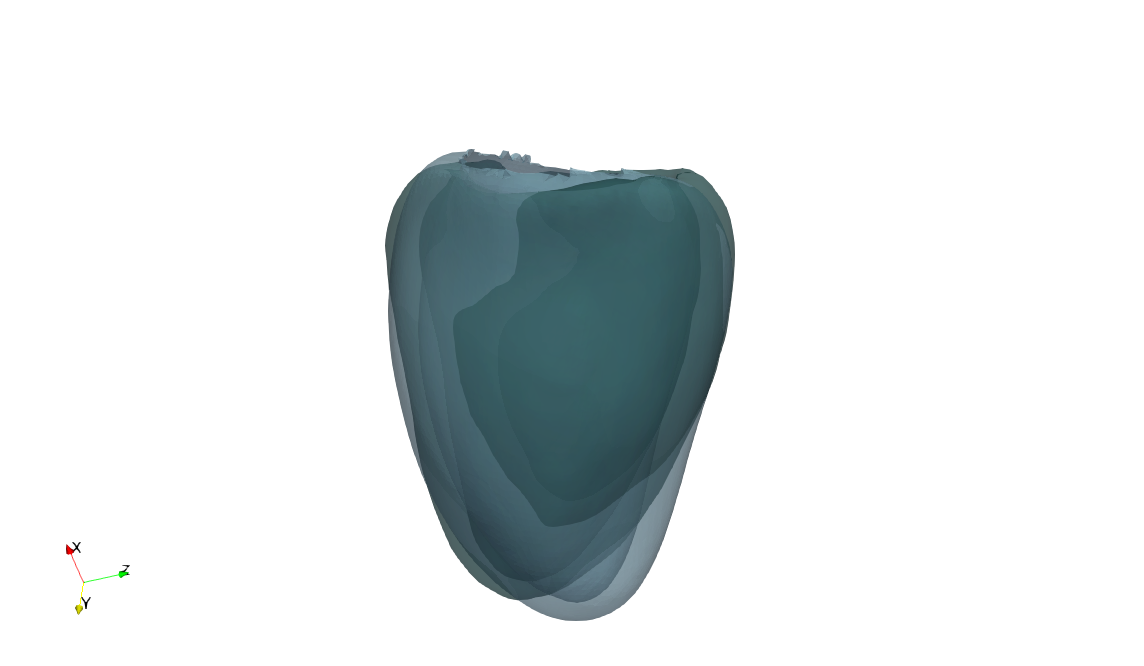} 
         \includegraphics[height=1.9cm, trim=13.5cm 1.5cm 14cm 3.5cm, clip]{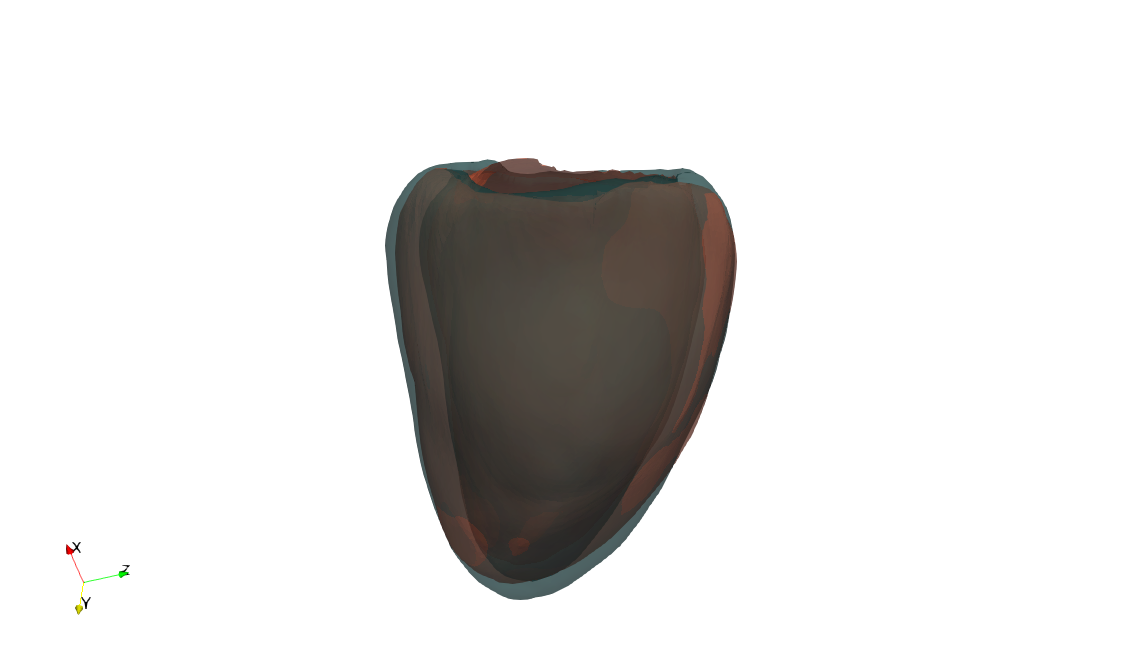}
    \end{tabular}
    }
    \hspace{-0.6cm}
    \setcounter{subfigure}{1}
    \subfloat[Segmentations]{
    \begin{tabular}{c}
         \includegraphics[height=1.9cm]{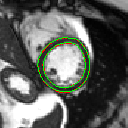} 
         \includegraphics[height=1.9cm]{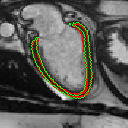} 
         \includegraphics[height=1.9cm]{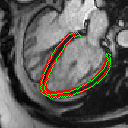} 
    \end{tabular}
    } 
    \hspace{-0.8cm}
    \caption{An example of ED frame mesh reconstruction. (a) Left: ground truth mesh (Green) of a subject heart vs. the template (Blue); right: ground truth mesh (Green) vs. the reconstructed mesh (Red). (b) 2D contours on SAX, 2CH and 4CH view planes generated by rasterizing the reconstructed mesh on corresponding view planes. Red contours denote predicted results, while green contours denote ground truth.}
    \label{meshRecon}
\end{figure}

The proposed method first reconstructs the mesh of the ED frame for each test subject. Fig.~\ref{meshRecon} (a) shows that the reconstructed mesh fits the ground truth mesh for a sample case. We extracted SAX, 2CH and 4CH view planes from the reconstructed ED frame mesh and generated 2D segmentations on different view planes. Fig.~\ref{meshRecon} (b) and Table~\ref{meshRecon_quantitative} qualitatively and quantitatively show the effectiveness of the mesh reconstruction by comparing the generated and the ground truth 2D myocardial contours.  

\begin{table}[tb]
\centering
\caption{Mesh reconstruction performance by comparing the predicted and ground truth 2D myocardium contours on different view planes. The results are reported as ``mean (standard deviation)".}
\label{meshRecon_quantitative}
\resizebox{0.48\textwidth}{!}{
\begin{tabular}{l|ccc}
\toprule[1.2pt]
~~~~~ &
SAX &
2CH &
4CH \\
\midrule
HD (mm)                 & 
5.96 (2.14)          &
5.97 (1.83)          &
6.06 (1.94)          \\
BoundF ($\%$)                         & 
84.61 (5.97) &
90.32 (4.94) &
90.11 (3.97) \\
\bottomrule[1.2pt]
\end{tabular}
}
\end{table}

\subsubsection{Mesh motion estimation performance}
\begin{figure}[tb]
 \centering
 \begin{tabular}{@{\hspace{-1\tabcolsep}}c@{\hspace{0.1\tabcolsep}}c@{\hspace{0.1\tabcolsep}}c@{\hspace{0.1\tabcolsep}}c@{\hspace{0.1\tabcolsep}}c@{\hspace{0.1\tabcolsep}}c}
  \raisebox{1.5\height}{\rotatebox[origin=c]{90}{\makecell{~\scalebox{0.8}{\textbf{Mesh}}}}} &
  \includegraphics[height=1.7cm, trim=14cm 2cm 14cm 3.5cm, clip]{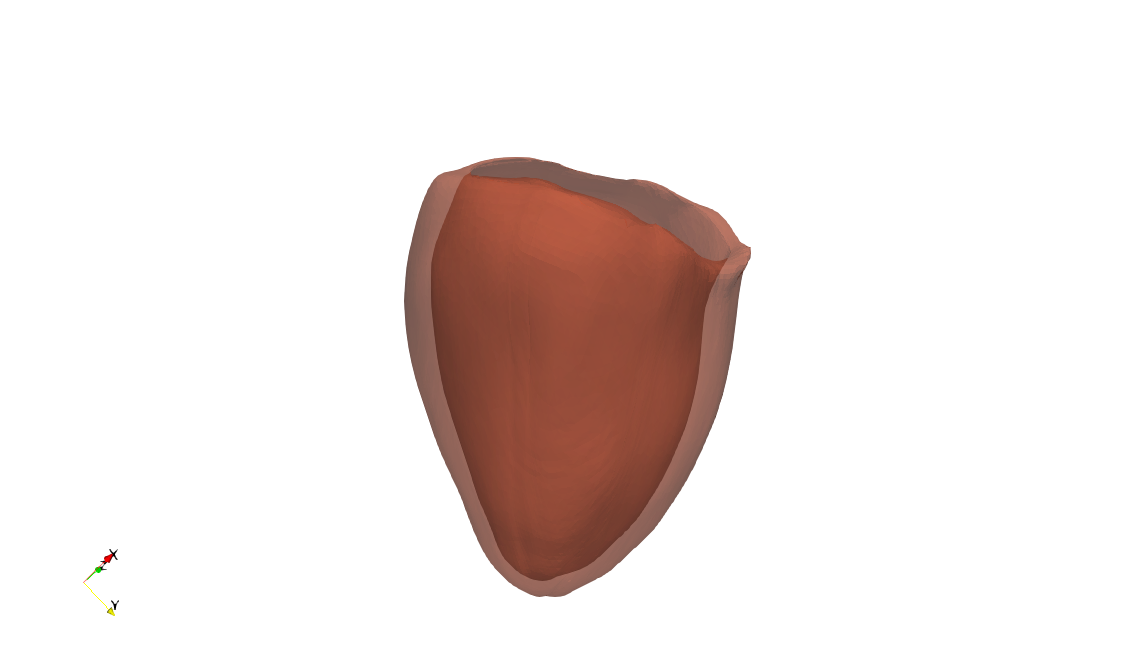} &
  \includegraphics[height=1.7cm, trim=14cm 2cm 14cm 3.5cm, clip]{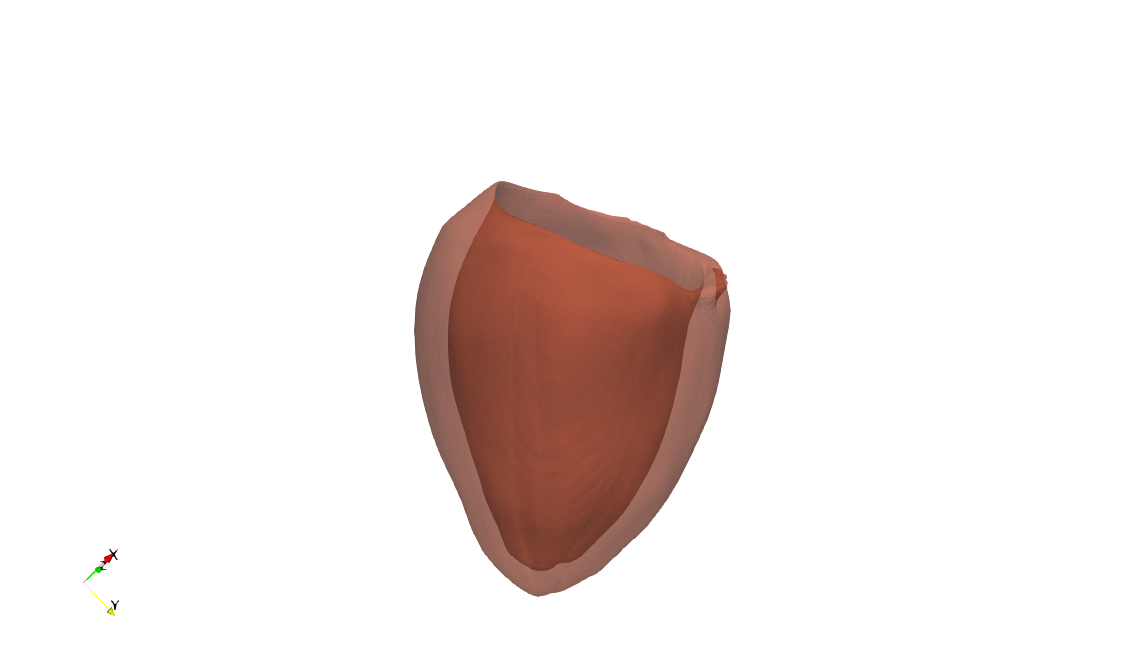} &
  \includegraphics[height=1.7cm, trim=14cm 2cm 14cm 3.5cm, clip]{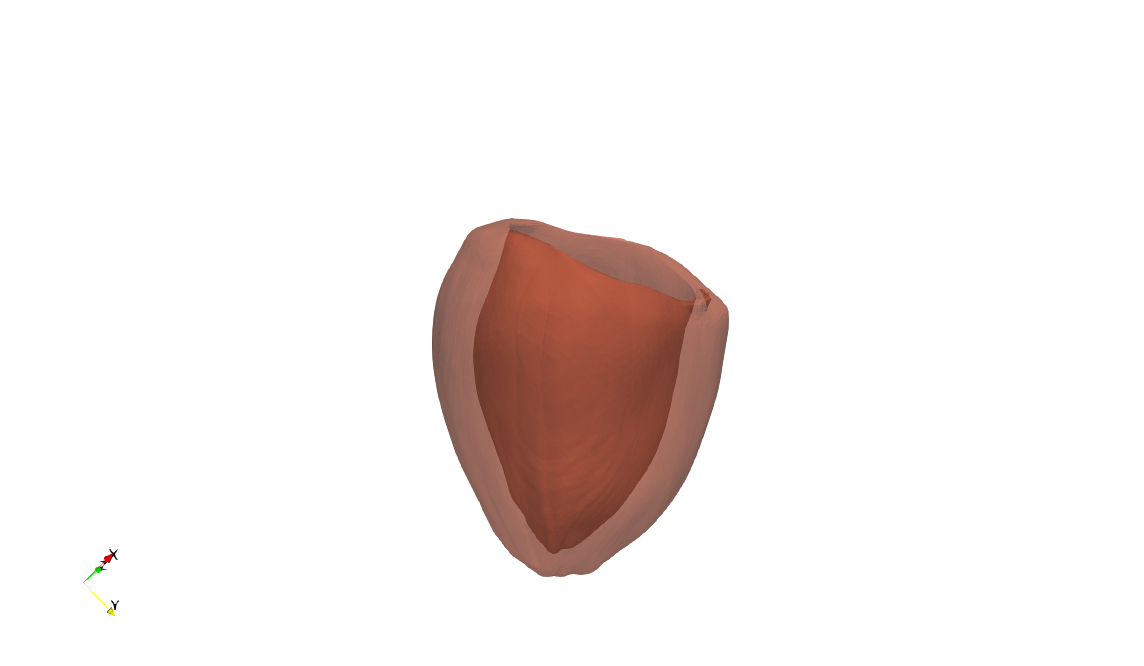} &
  \includegraphics[height=1.7cm, trim=14cm 2cm 14cm 3.5cm, clip]{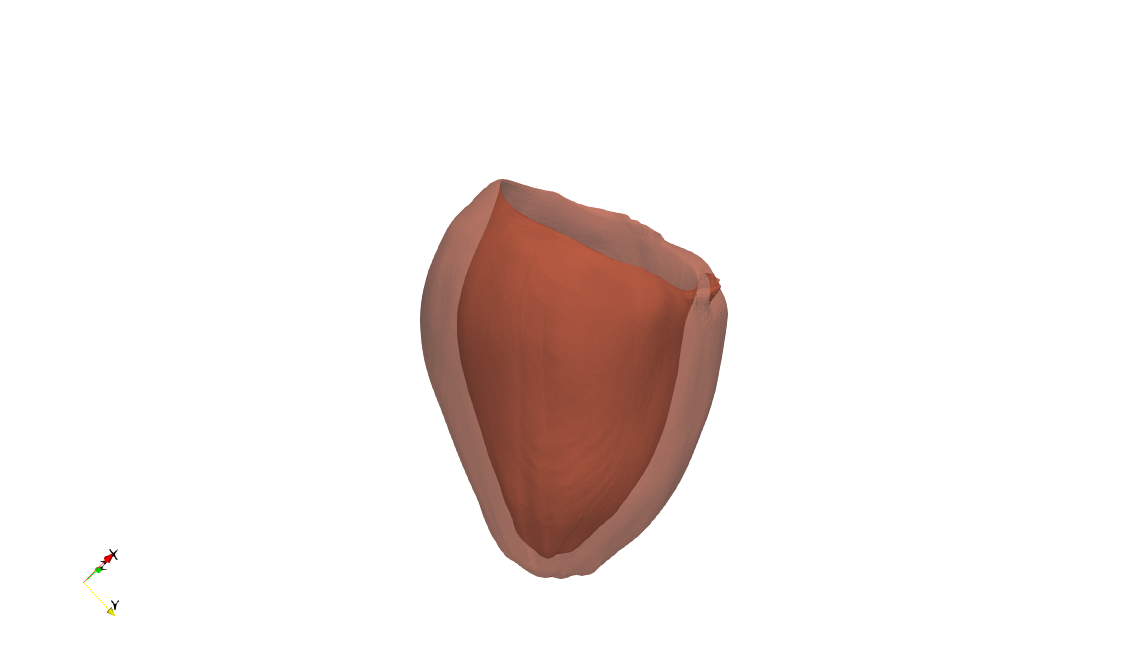} &
  \includegraphics[height=1.7cm, trim=14cm 2cm 14cm 3.5cm, clip]{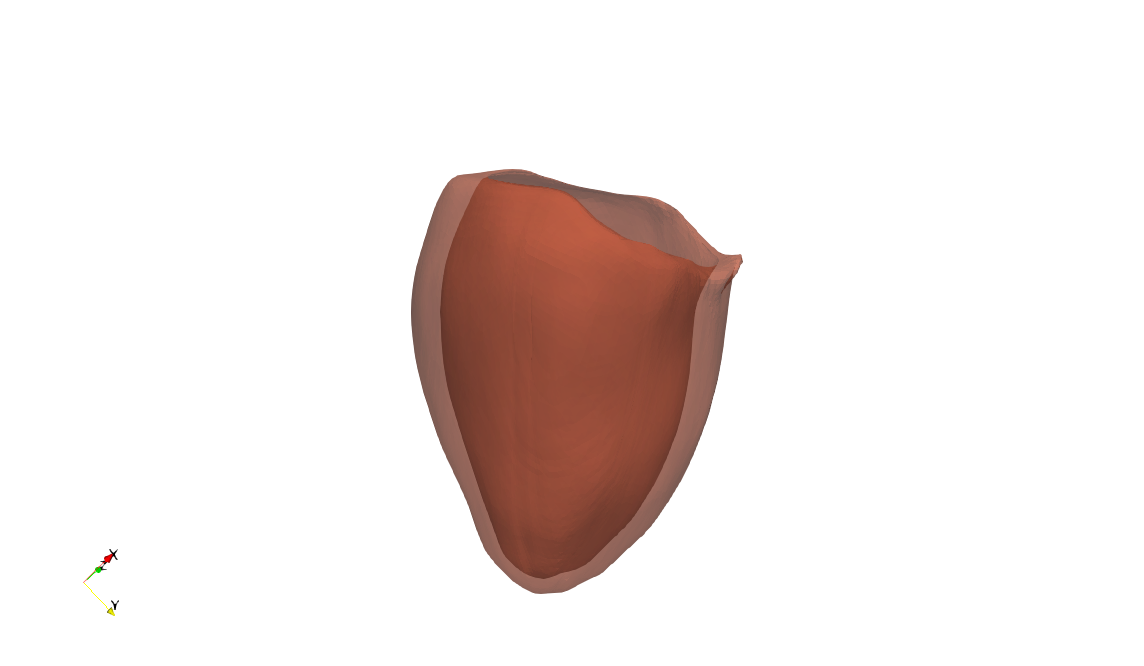} \\
  \raisebox{2\height}{\rotatebox[origin=c]{90}{\makecell{~\scalebox{0.8}{\textbf{SAX}}}}} &
  \includegraphics[height=1.7cm]{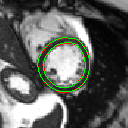} &
  \includegraphics[height=1.7cm]{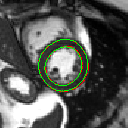} &
  \includegraphics[height=1.7cm]{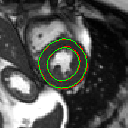} &
  \includegraphics[height=1.7cm]{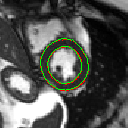} &
  \includegraphics[height=1.7cm]{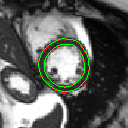} \\
  \raisebox{2\height}{\rotatebox[origin=c]{90}{\makecell{~\scalebox{0.8}{\textbf{2CH}}}}} &
  \includegraphics[height=1.7cm]{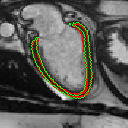} &
  \includegraphics[height=1.7cm]{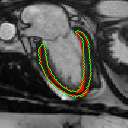} &
  \includegraphics[height=1.7cm]{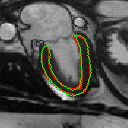} &
  \includegraphics[height=1.7cm]{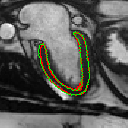} &
  \includegraphics[height=1.7cm]{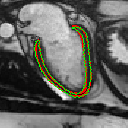} \\
  \raisebox{2\height}{\rotatebox[origin=c]{90}{\makecell{~\scalebox{0.8}{\textbf{4CH}}}}} &
  \includegraphics[height=1.7cm]{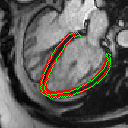} &
  \includegraphics[height=1.7cm]{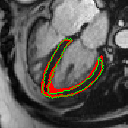} &
  \includegraphics[height=1.7cm]{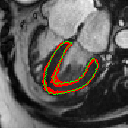} &
  \includegraphics[height=1.7cm]{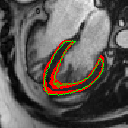} &
  \includegraphics[height=1.7cm]{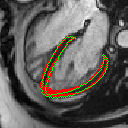} \\
  ~~~ &
  \raisebox{0.1\height}{\rotatebox[origin=c]{0}{\makecell{~\scalebox{0.8}{\textbf{t=0}}}}} &
  \raisebox{0.1\height}{\rotatebox[origin=c]{0}{\makecell{~\scalebox{0.8}{\textbf{t=10}}}}} &
  \raisebox{0.1\height}{\rotatebox[origin=c]{0}{\makecell{~\scalebox{0.8}{\textbf{t=20}}}}} &
  \raisebox{0.1\height}{\rotatebox[origin=c]{0}{\makecell{~\scalebox{0.8}{\textbf{t=30}}}}} &
  \raisebox{0.1\height}{\rotatebox[origin=c]{0}{\makecell{~\scalebox{0.8}{\textbf{t=40}}}}}
  \end{tabular}
  \caption{Examples of motion tracking results. The reconstructed ED frame mesh is deformed to the $t$-th frame using the estimated 3D mesh motion fields. 2D myocardium contours on SAX, 2CH and 4CH view planes (Row 2-4) are generated by extracting the corresponding planes from the predicted $t$-th frame mesh. Red contours are predicted results while green contours are ground truth.}
  \label{singlesub_results}
\end{figure}

Following mesh reconstruction, the proposed method estimates mesh motion fields in the full cardiac cycle. For each test subject, with the obtained vertex-wise motion fields $\{\Delta V_t|t=[0,49]\}$, the reconstructed ED frame mesh is deformed to the $t$-th frame. Red meshes in Fig.~\ref{singlesub_results} shows that the estimated mesh motion field $\Delta V_{0\rightarrow t}$ enables 3D myocardial motion tracking on meshes. In addition, we extracted SAX/2CH/4CH view planes from the predicted $t$-th frame mesh and generated the predicted 2D myocardium contours on different view planes. Fig.~\ref{singlesub_results} shows the effectiveness of $\Delta V_{0\rightarrow t}$ by comparing the predicted and the ground truth 2D myocardium contours. 
 
\subsubsection{Comparison study}
\begin{figure}[tb]
 \centering
 \begin{tabular}{@{\hspace{-1.5\tabcolsep}}c@{\hspace{0.7\tabcolsep}}c@{\hspace{0.1\tabcolsep}}c@{\hspace{0.1\tabcolsep}}c@{\hspace{0.1\tabcolsep}}c@{\hspace{0.1\tabcolsep}}c@{\hspace{-1\tabcolsep}}}
  \raisebox{1.7\height}{\rotatebox[origin=c]{90}{\makecell{~\scalebox{0.8}{\textbf{FFD}}\\~\scalebox{0.8}{\textbf{\cite{Rueckert1999}}}}}} &
  \includegraphics[height=1.7cm, trim=14cm 2cm 14cm 3.5cm, clip]{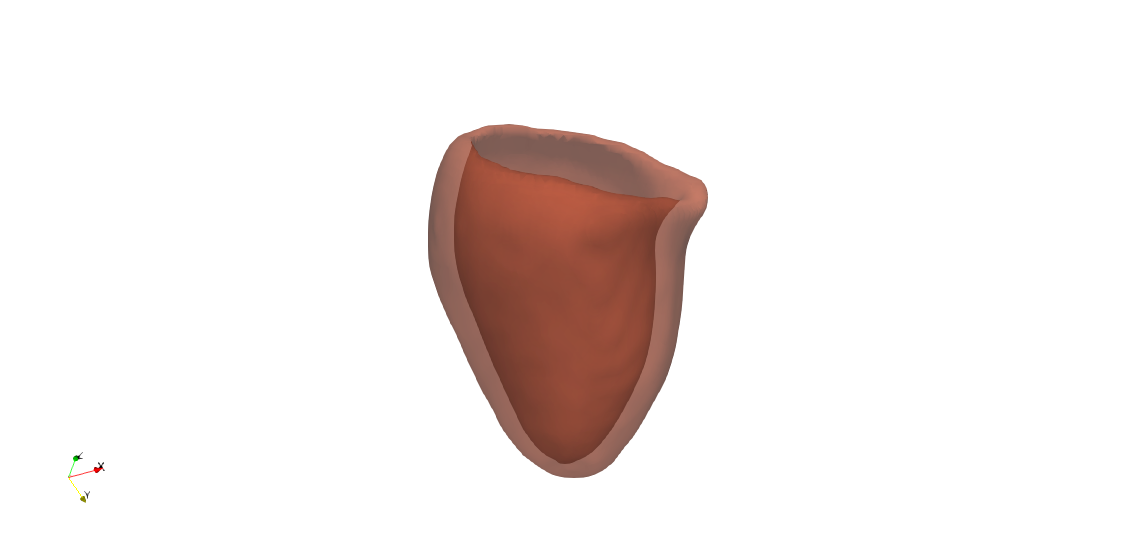} &
  \includegraphics[height=1.7cm, trim=14cm 2cm 14cm 3.5cm, clip]{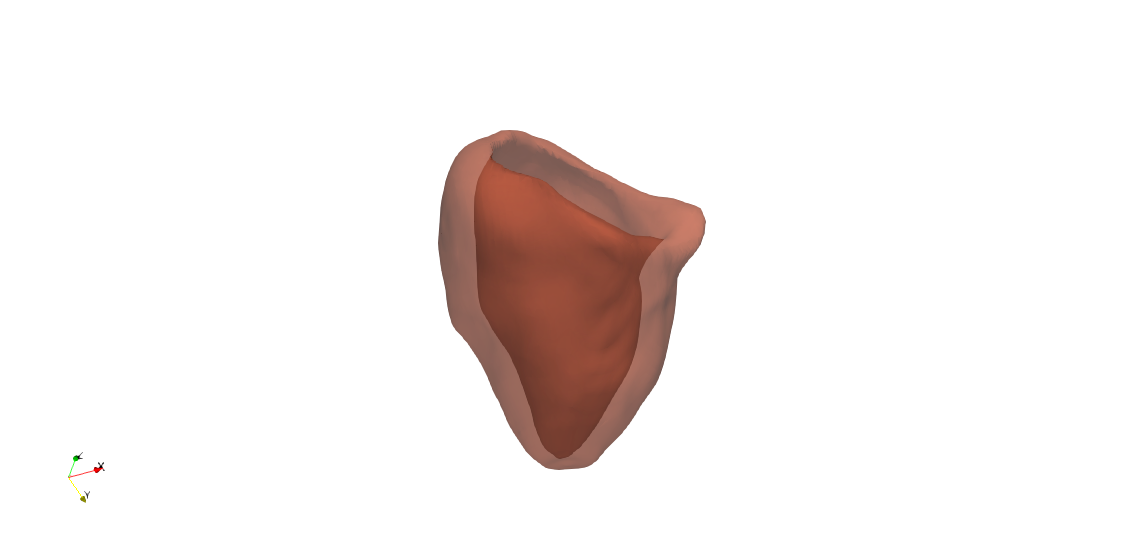} &
  \includegraphics[height=1.7cm, trim=14cm 2cm 14cm 3.5cm, clip]{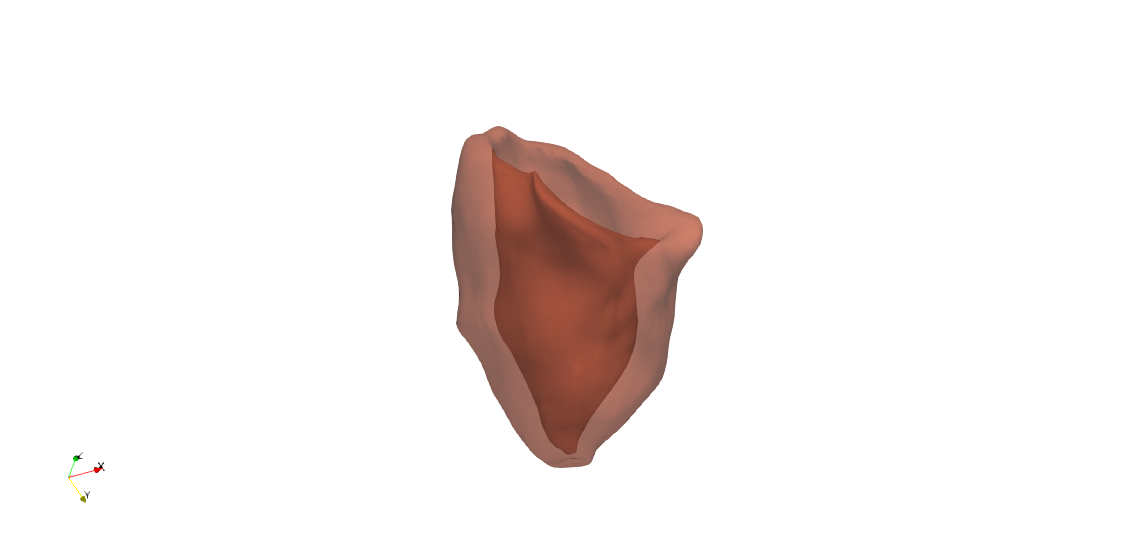} &
  \includegraphics[height=1.7cm, trim=14cm 2cm 14cm 3.5cm, clip]{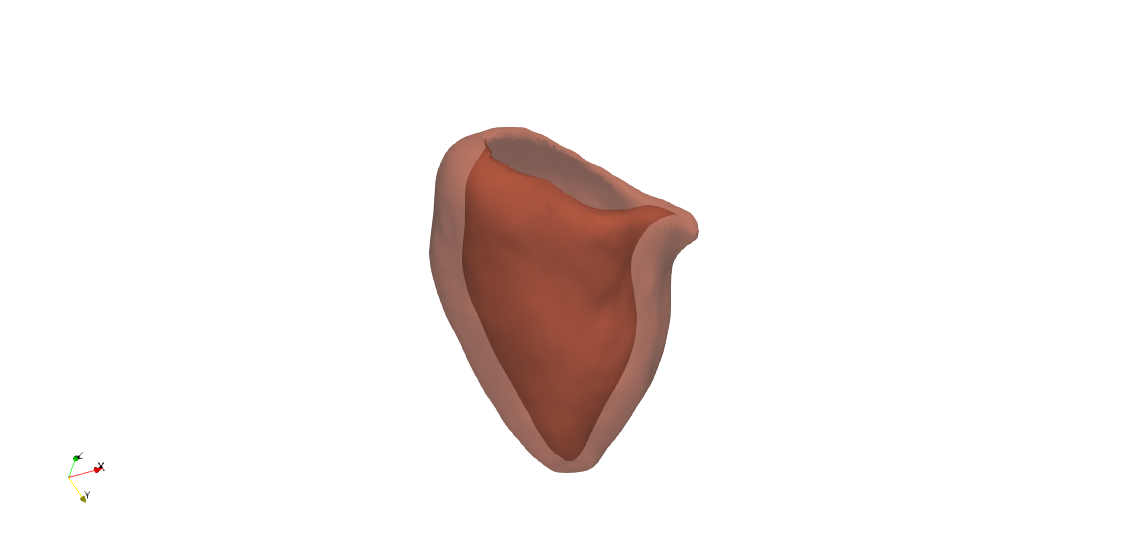} &
  \includegraphics[height=1.7cm, trim=14cm 2cm 14cm 3.5cm, clip]{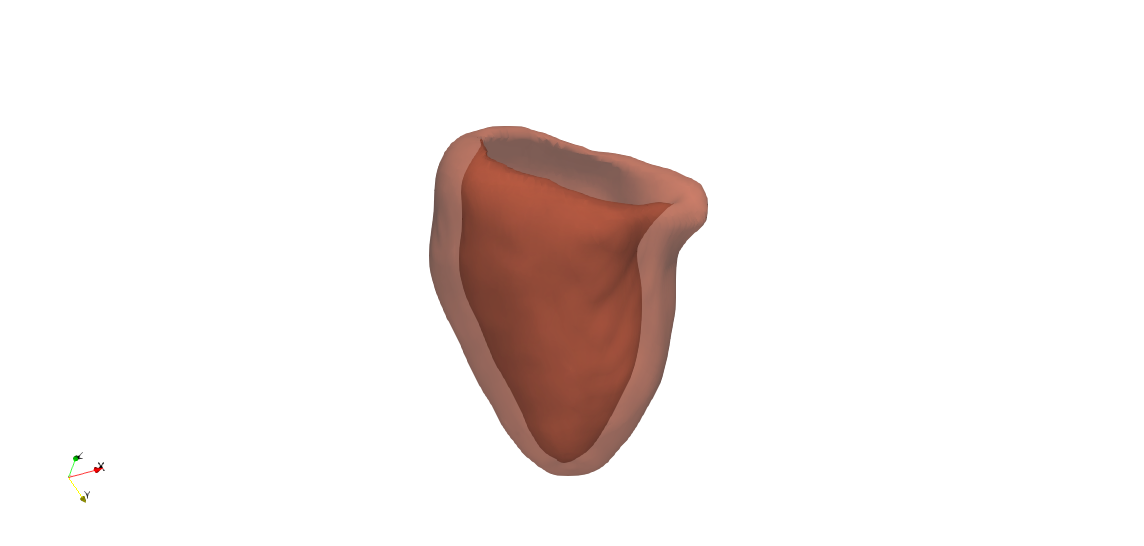} \\
  \raisebox{1.2\height}{\rotatebox[origin=c]{90}{\makecell{~\scalebox{0.8}{\textbf{dDemons}}\\~\scalebox{0.8}{\textbf{\cite{Vercauteren2007}}}}}} &
  \includegraphics[height=1.7cm, trim=14cm 2cm 14cm 3.5cm, clip]{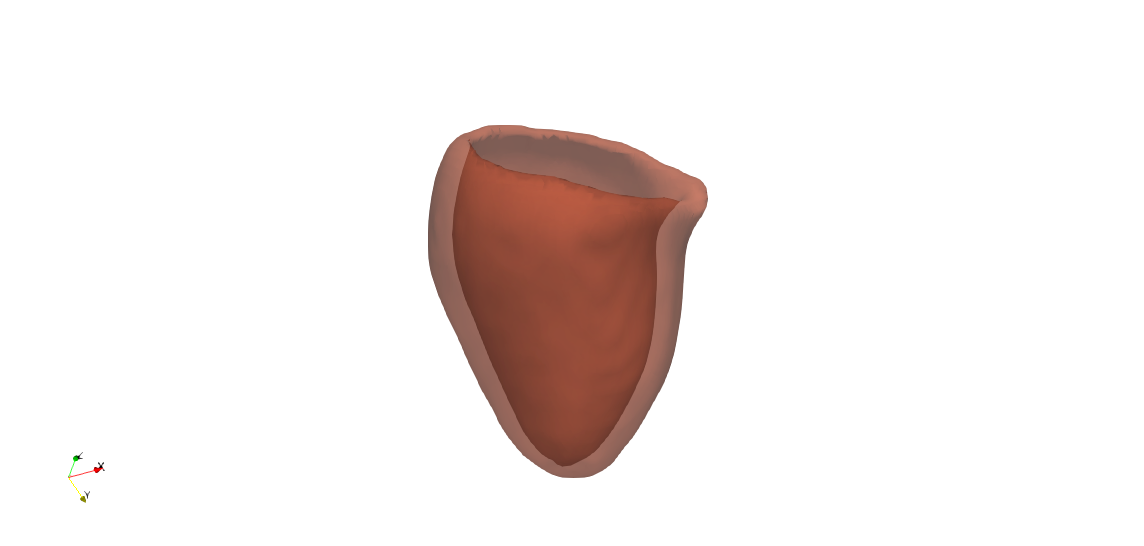} &
  \includegraphics[height=1.7cm, trim=14cm 2cm 14cm 3.5cm, clip]{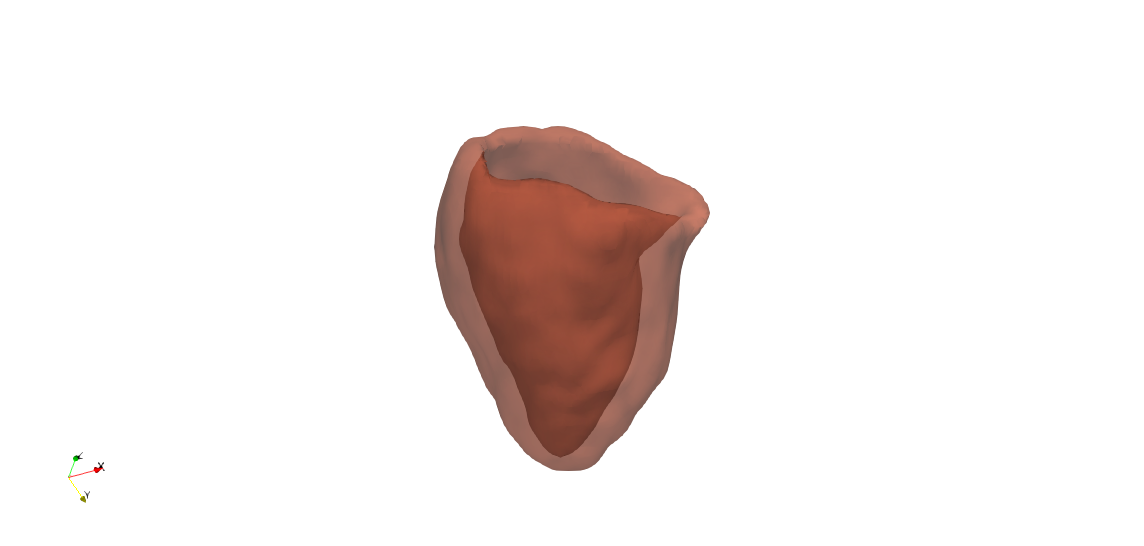} &
  \includegraphics[height=1.7cm, trim=14cm 2cm 14cm 3.5cm, clip]{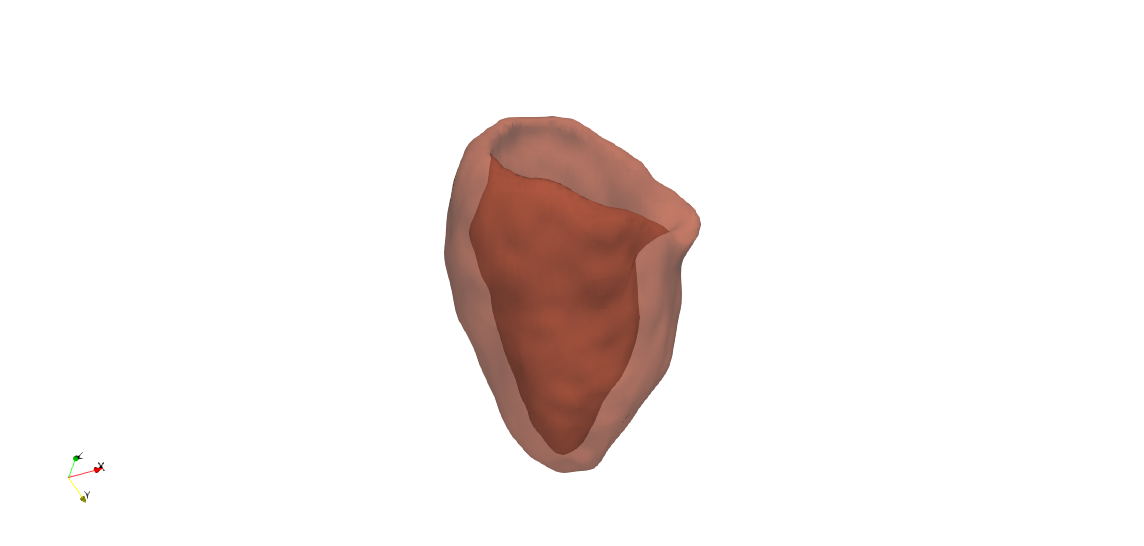} &
  \includegraphics[height=1.7cm, trim=14cm 2cm 14cm 3.5cm, clip]{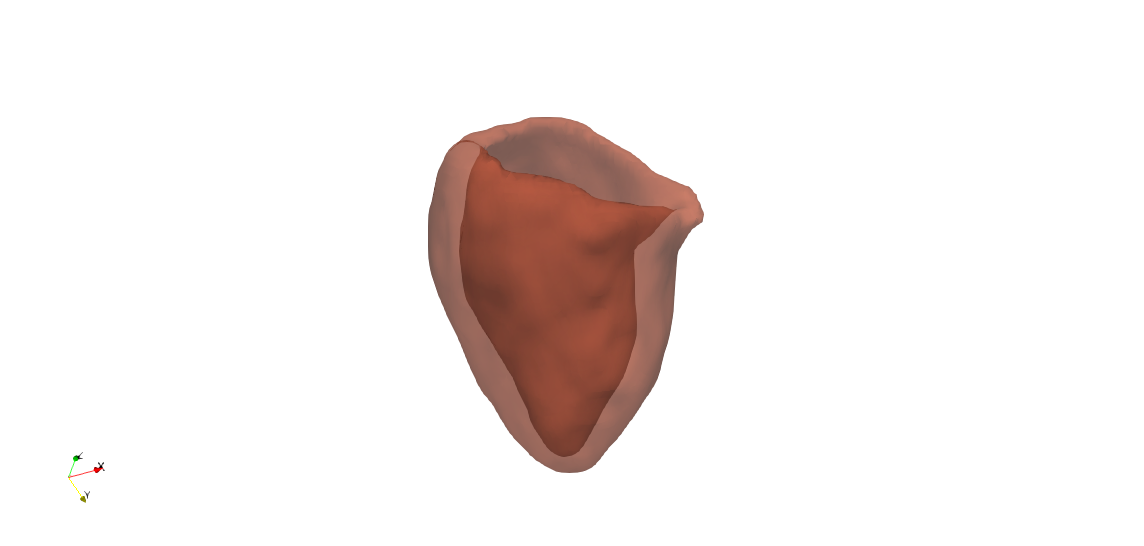} &
  \includegraphics[height=1.7cm, trim=14cm 2cm 14cm 3.5cm, clip]{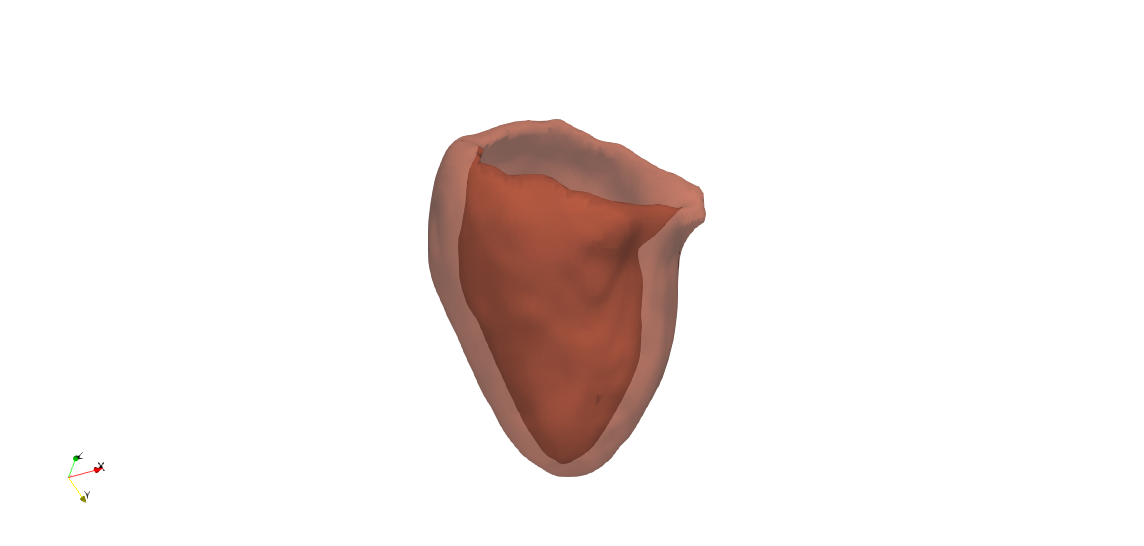} \\
  \raisebox{1.2\height}{\rotatebox[origin=c]{90}{\makecell{~\scalebox{0.8}{\textbf{3D-UNet}}\\~\scalebox{0.8}{\textbf{\cite{ociek2016}}}}}} &
  \includegraphics[height=1.7cm, trim=14cm 2cm 14cm 3.5cm, clip]{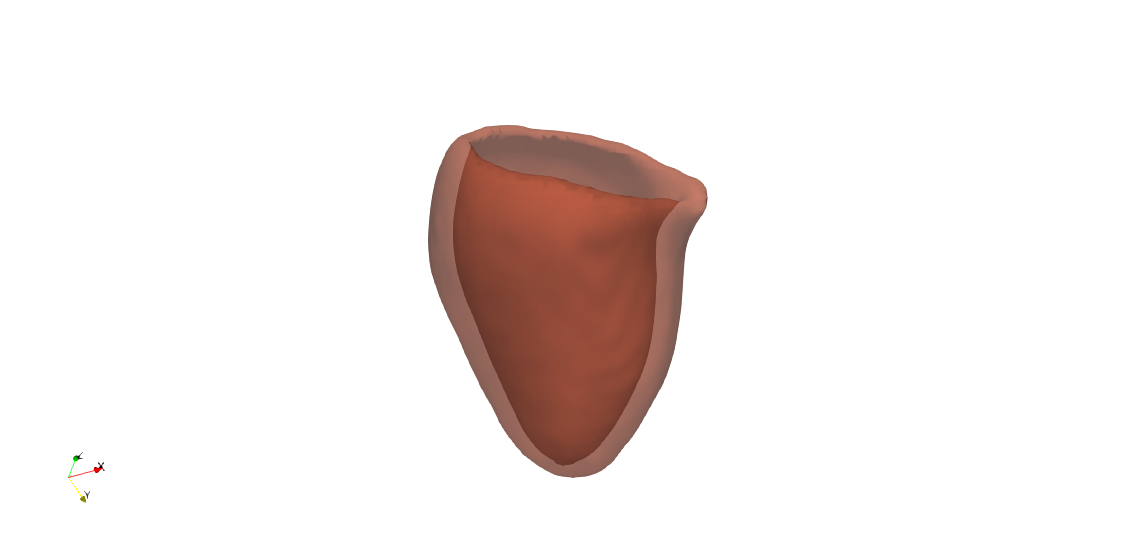} &
  \includegraphics[height=1.7cm, trim=14cm 2cm 14cm 3.5cm, clip]{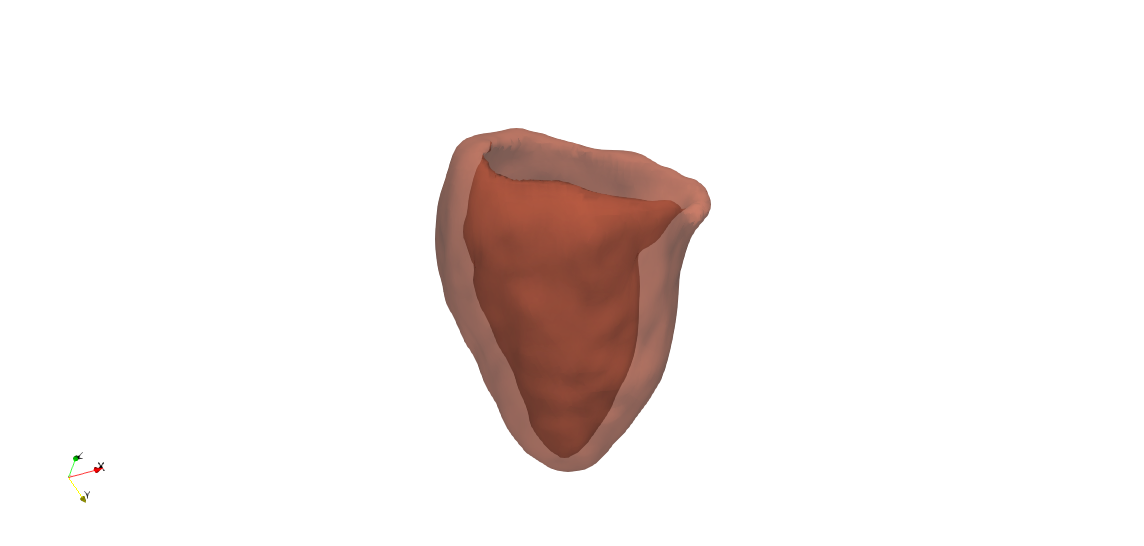} &
  \includegraphics[height=1.7cm, trim=14cm 2cm 14cm 3.5cm, clip]{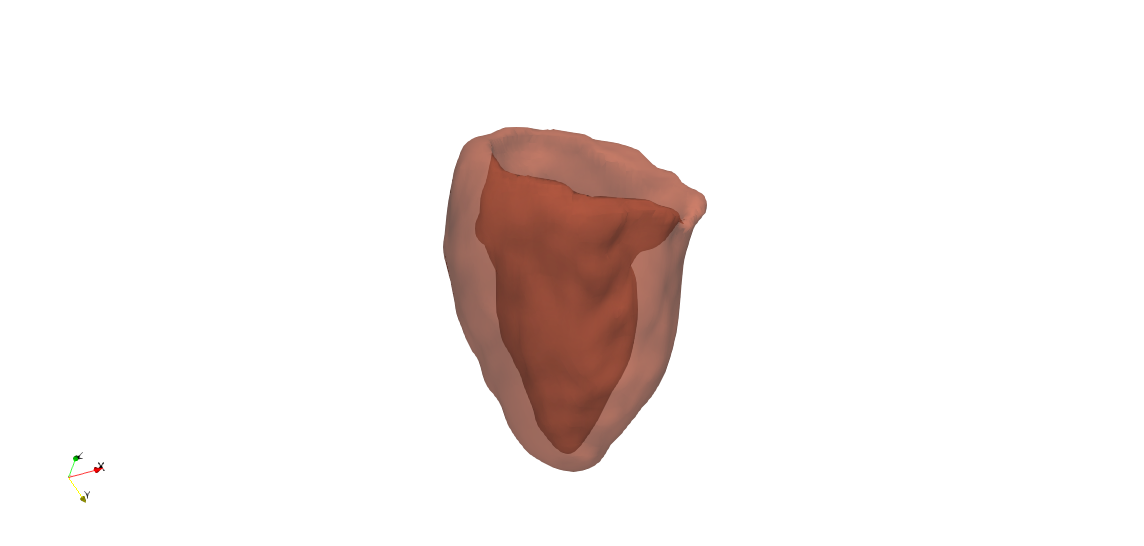} &
  \includegraphics[height=1.7cm, trim=14cm 2cm 14cm 3.5cm, clip]{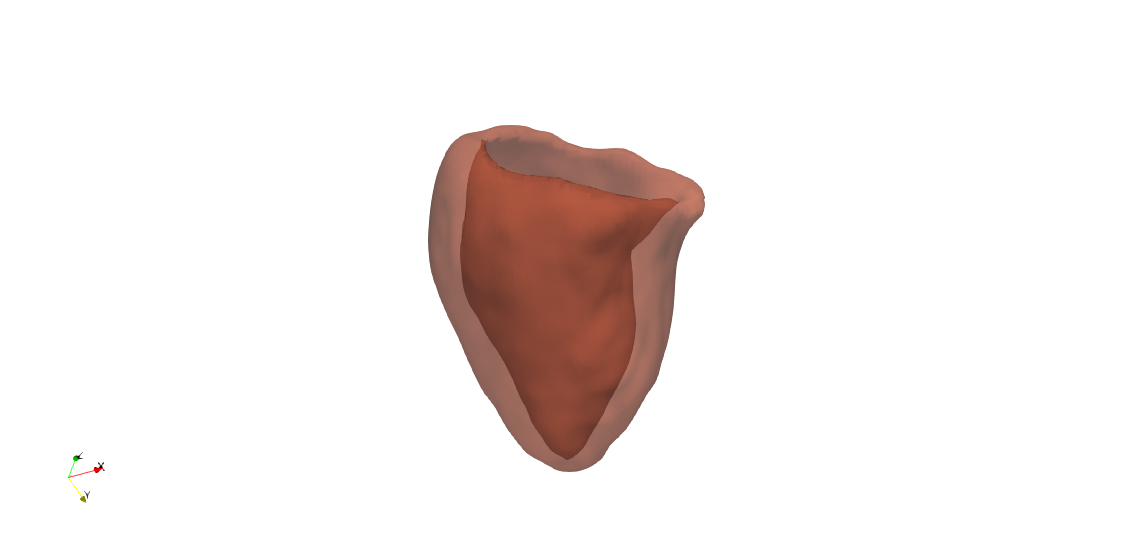} &
  \includegraphics[height=1.7cm, trim=14cm 2cm 14cm 3.5cm, clip]{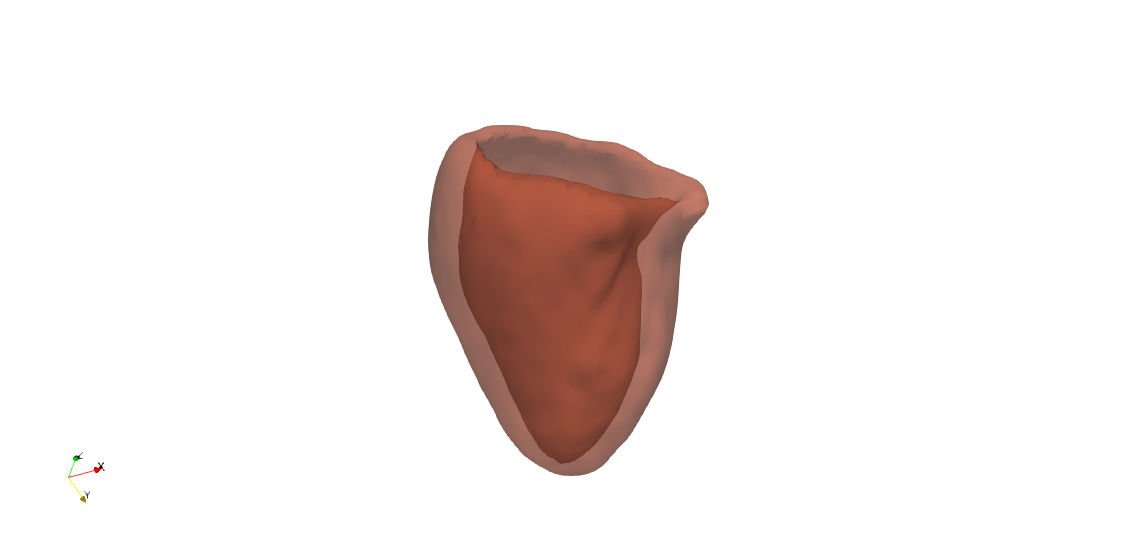} \\
  \raisebox{0.7\height}{\rotatebox[origin=c]{90}{\makecell{~\scalebox{0.8}{\textbf{MulViMotion}}\\~\scalebox{0.8}{\textbf{\cite{Meng2022_mulvimotion}}}}}} &
  \includegraphics[height=1.7cm, trim=14cm 2cm 14cm 3.5cm, clip]{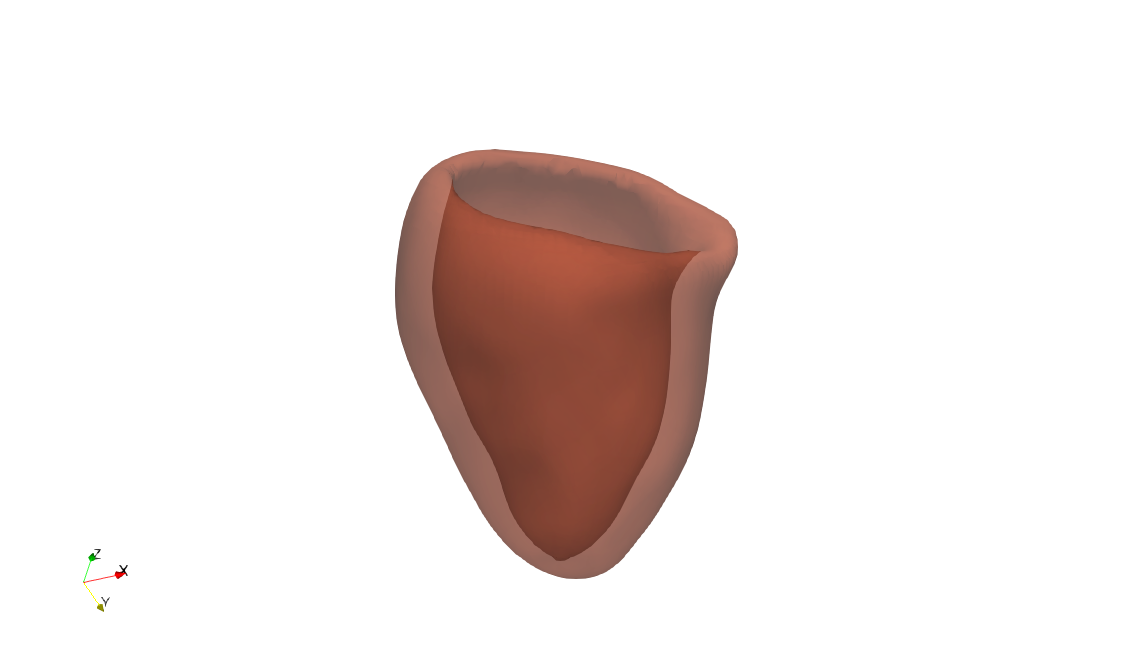} &
  \includegraphics[height=1.7cm, trim=14cm 2cm 14cm 3.5cm, clip]{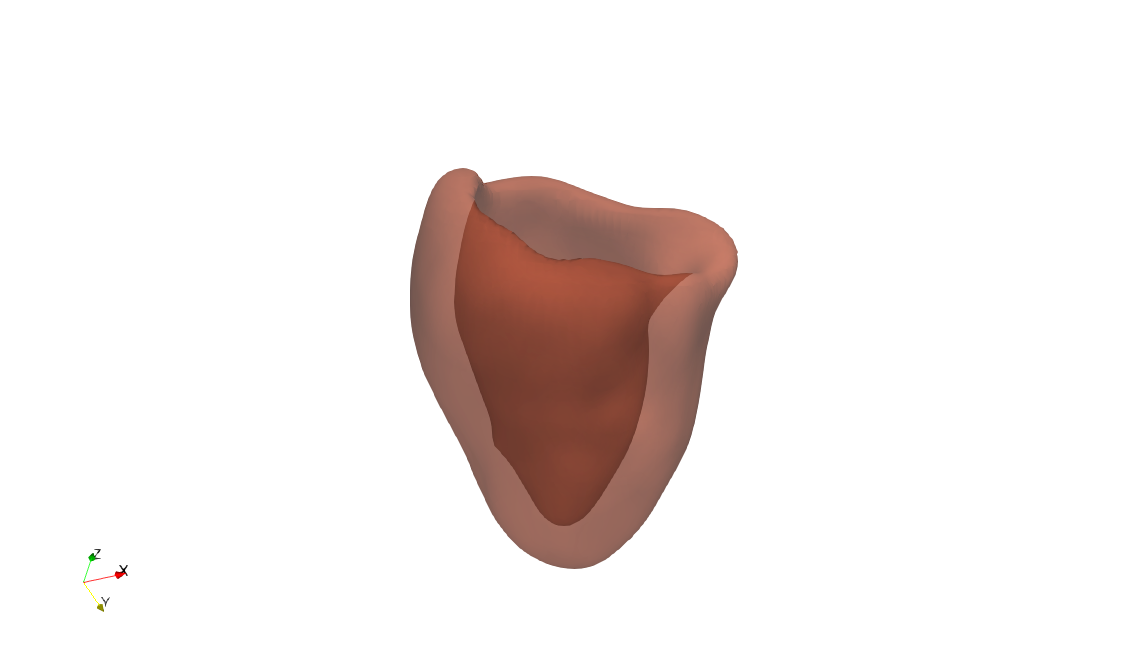} &
  \includegraphics[height=1.7cm, trim=14cm 2cm 14cm 3.5cm, clip]{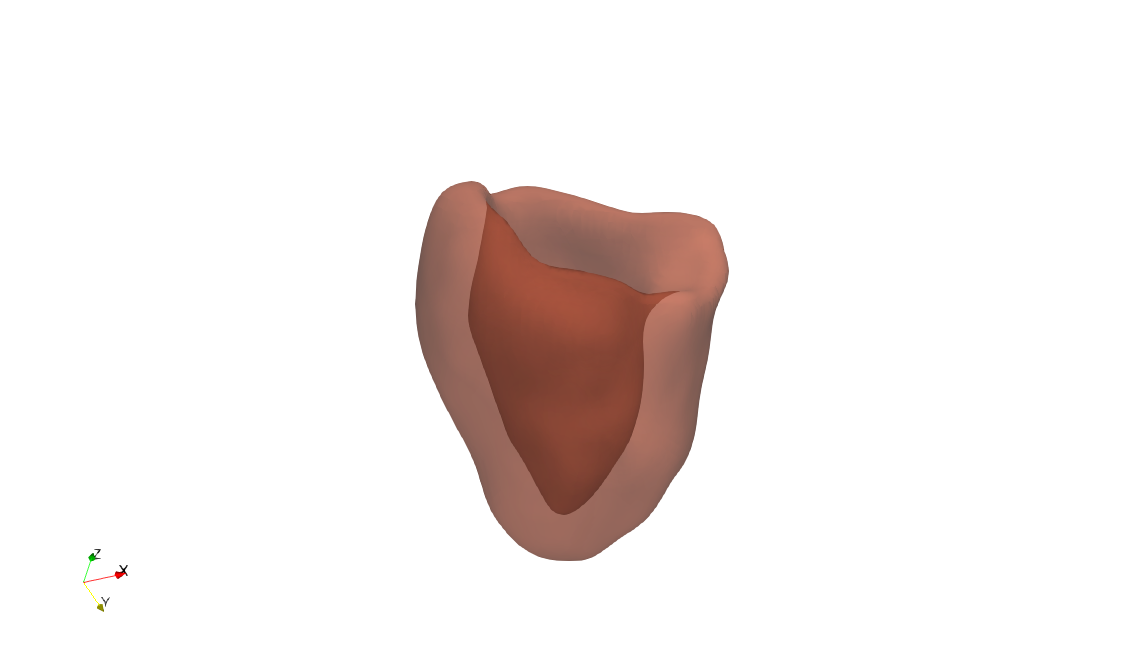} &
  \includegraphics[height=1.7cm, trim=14cm 2cm 14cm 3.5cm, clip]{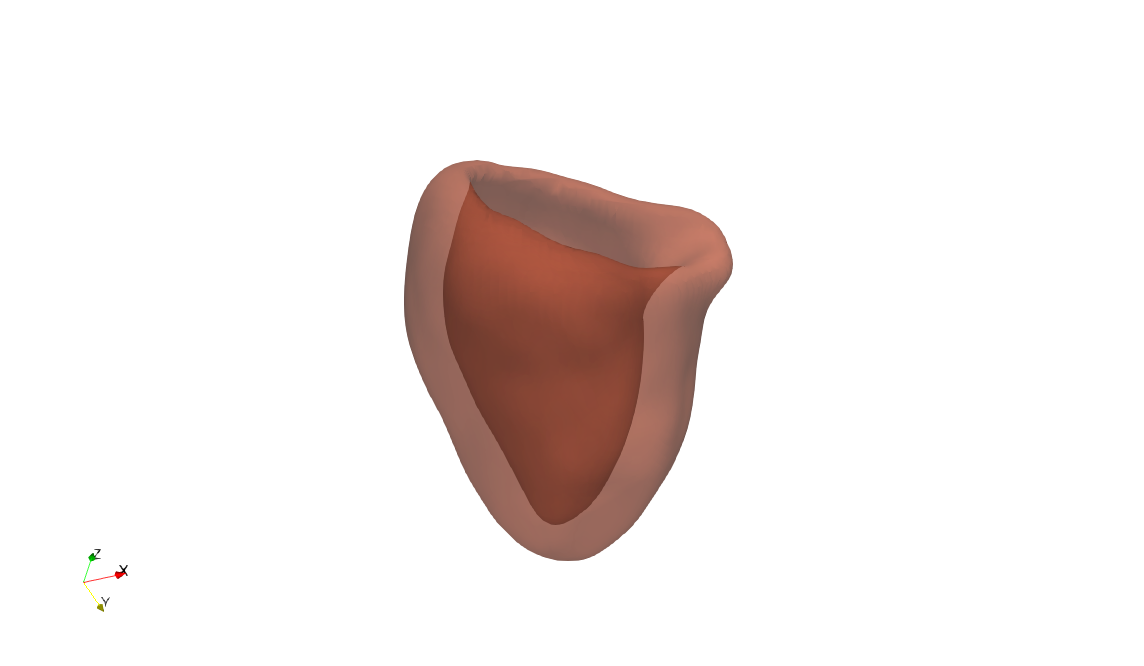} &
  \includegraphics[height=1.7cm, trim=14cm 2cm 14cm 3.5cm, clip]{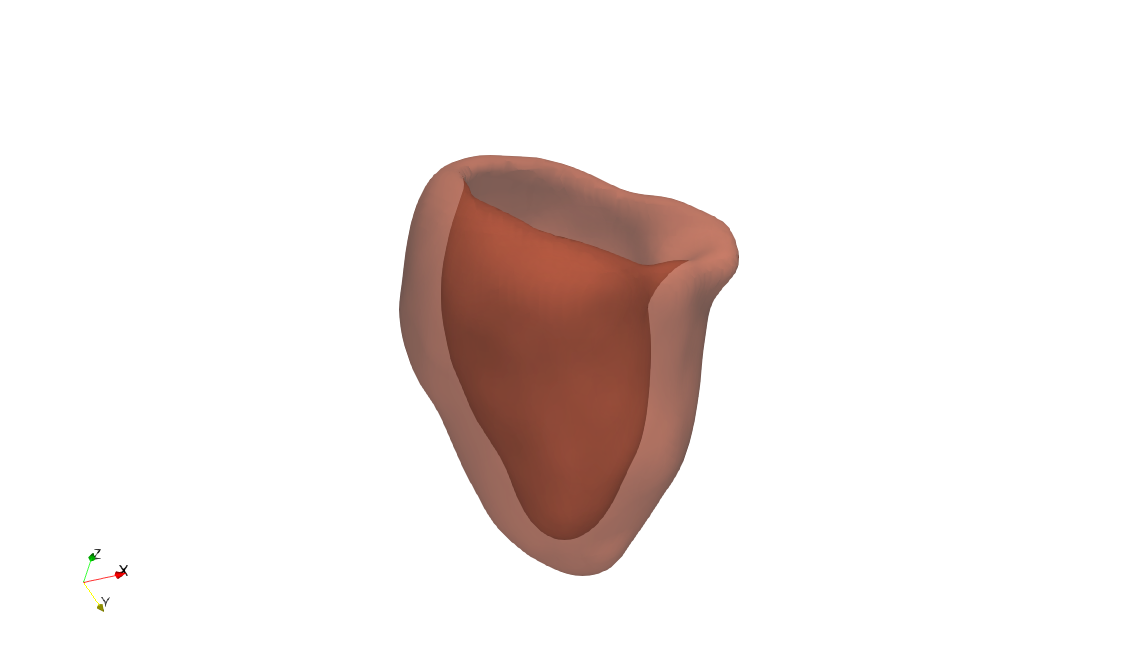} \\
  \raisebox{0.7\height}{\rotatebox[origin=c]{90}{\makecell{~\scalebox{0.8}{\textbf{MeshMotion}}\\~\scalebox{0.8}{\textbf{\cite{Meng2022_miccai}}}}}} &
  \includegraphics[height=1.7cm, trim=14cm 2cm 14cm 3.5cm, clip]{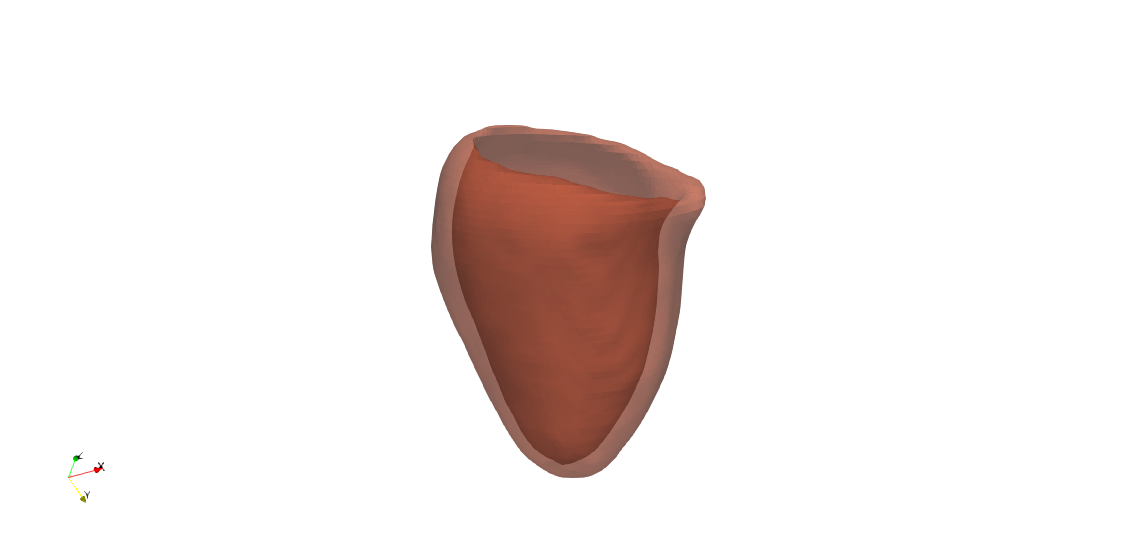} &
  \includegraphics[height=1.7cm, trim=14cm 2cm 14cm 3.5cm, clip]{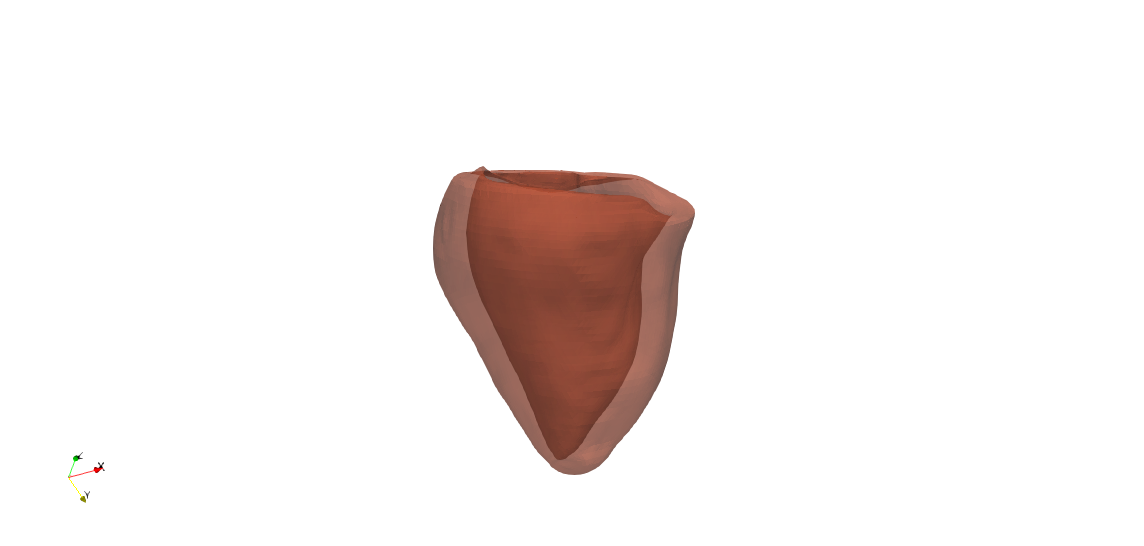} &
  \includegraphics[height=1.7cm, trim=14cm 2cm 14cm 3.5cm, clip]{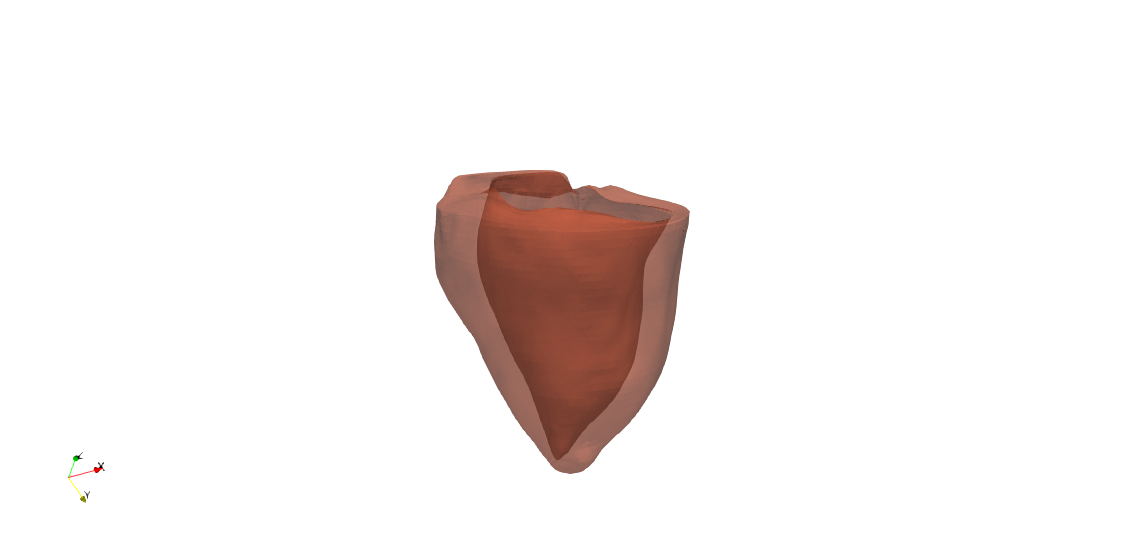} &
  \includegraphics[height=1.7cm, trim=14cm 2cm 14cm 3.5cm, clip]{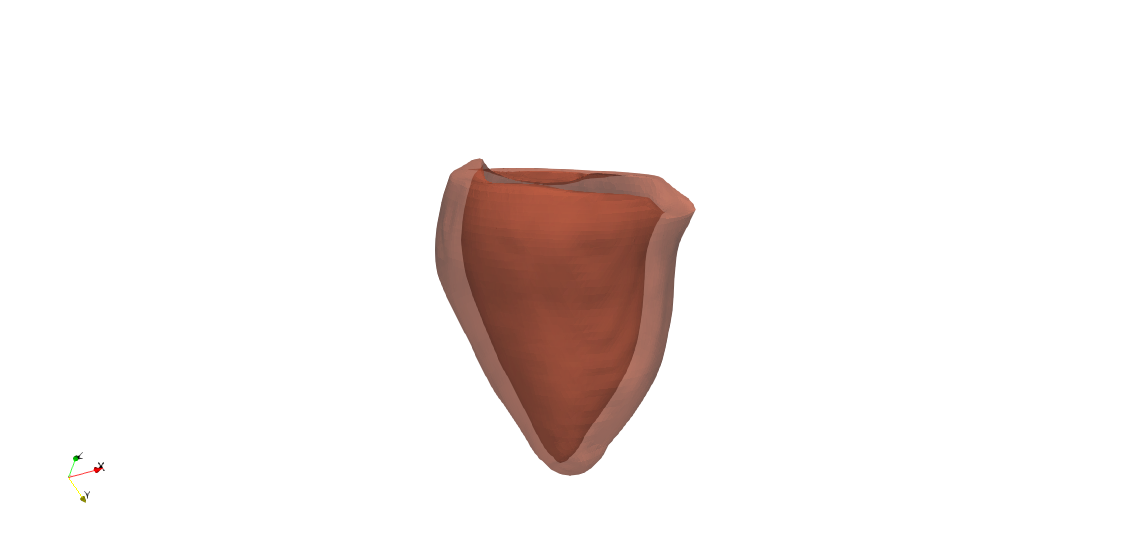} &
  \includegraphics[height=1.7cm, trim=14cm 2cm 14cm 3.5cm, clip]{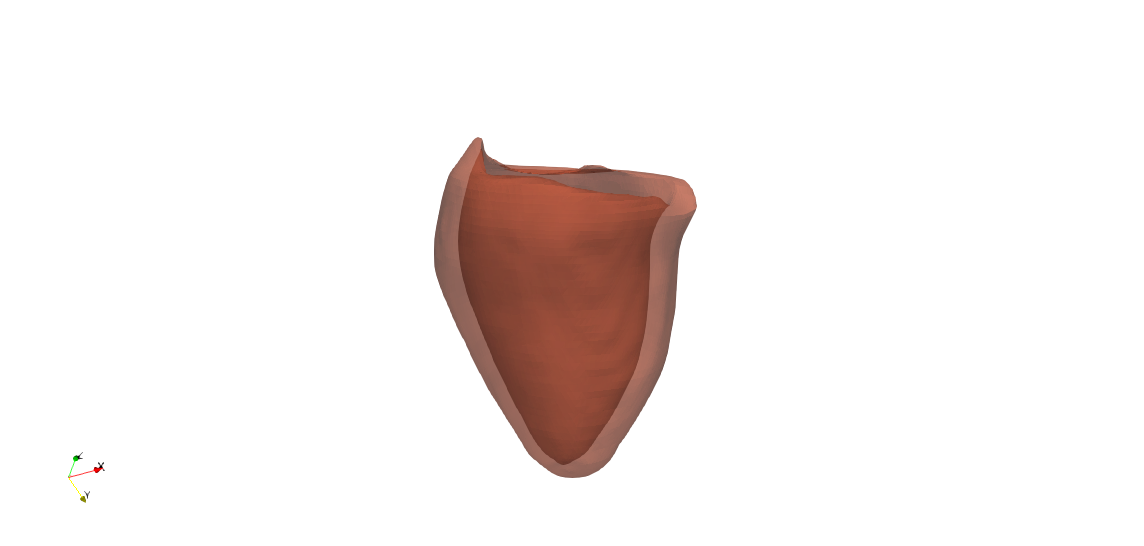} \\
  \raisebox{0.5\height}{\rotatebox[origin=c]{90}{\makecell{~\scalebox{0.8}{\textbf{DeepMesh}}}}} &
  \includegraphics[height=1.7cm, trim=14cm 2cm 14cm 3.5cm, clip]{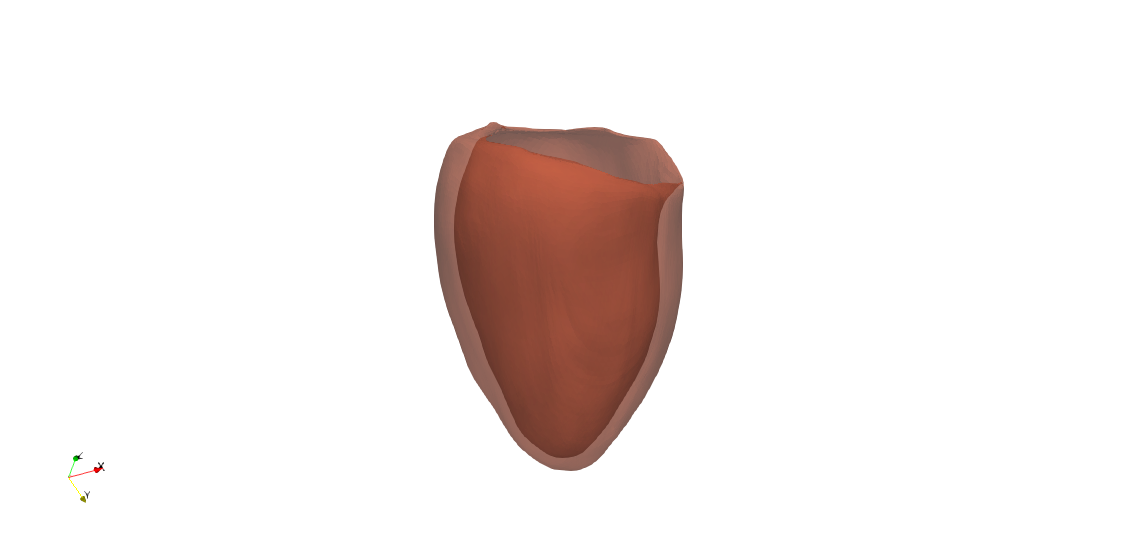} &
  \includegraphics[height=1.7cm, trim=14cm 2cm 14cm 3.5cm, clip]{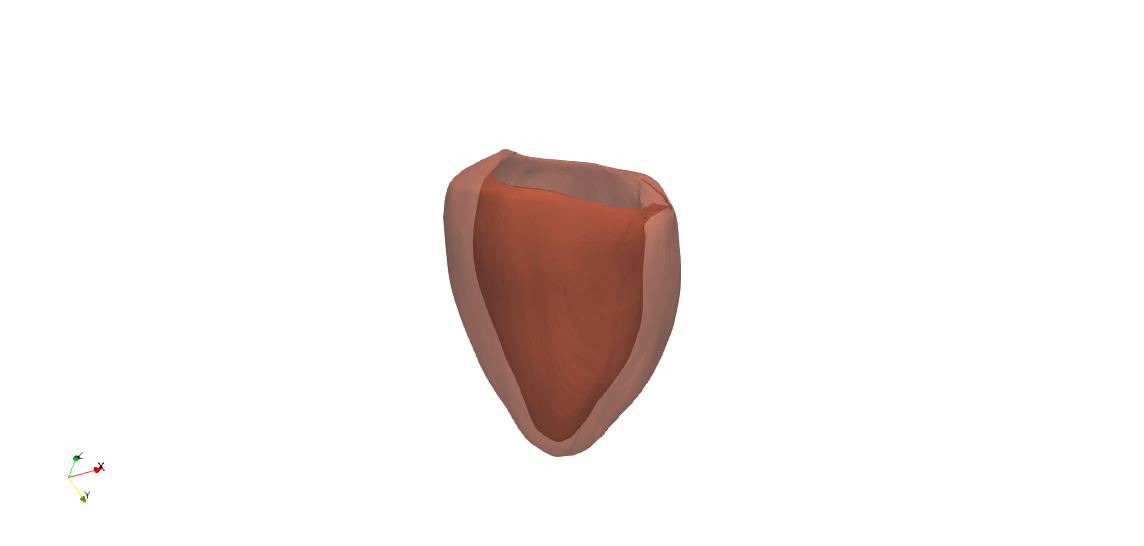} &
  \includegraphics[height=1.7cm, trim=14cm 2cm 14cm 3.5cm, clip]{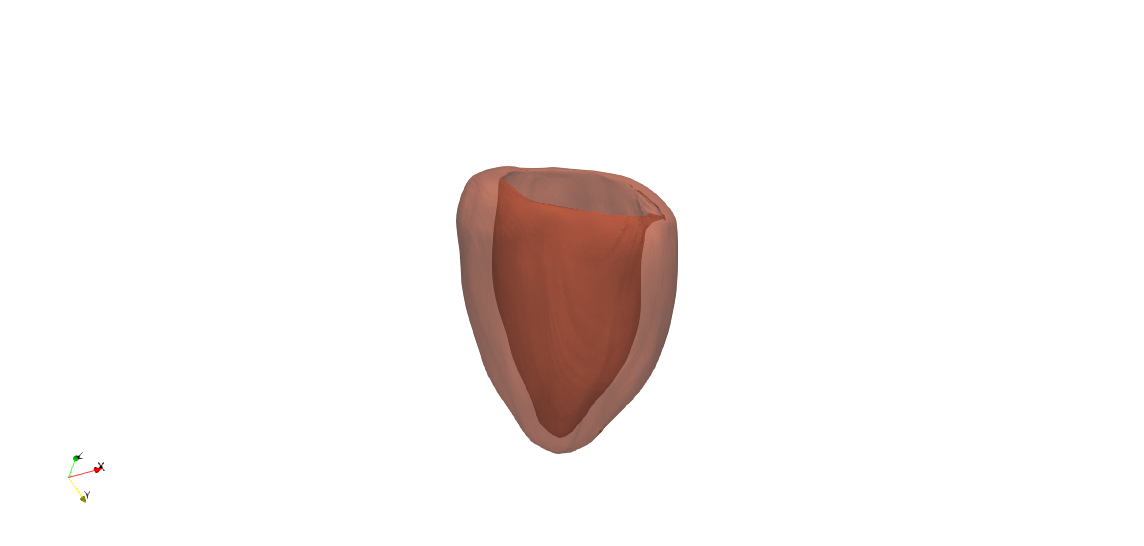} &
  \includegraphics[height=1.7cm, trim=14cm 2cm 14cm 3.5cm, clip]{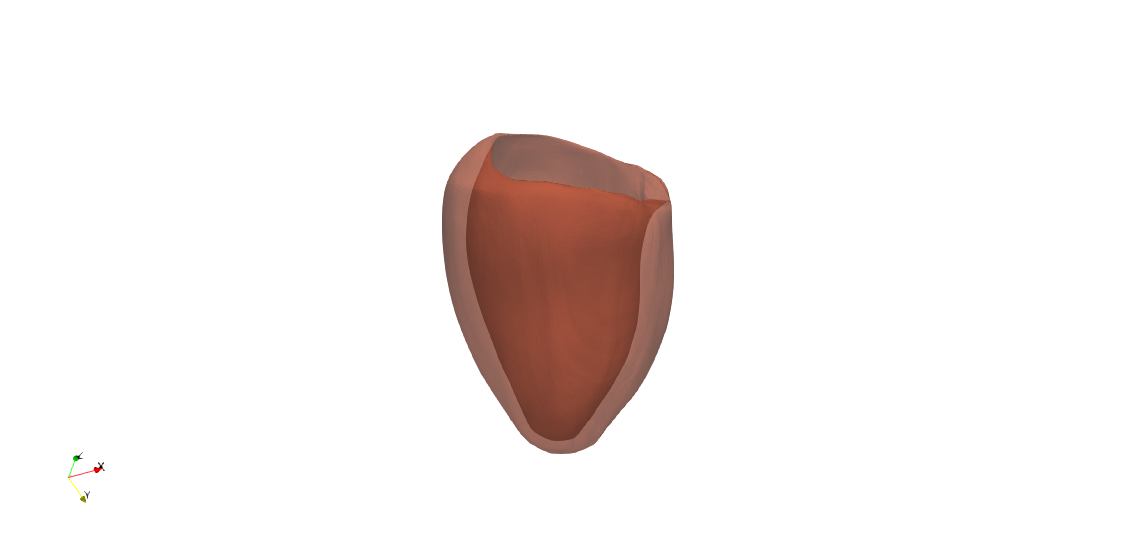} &
  \includegraphics[height=1.7cm, trim=14cm 2cm 14cm 3.5cm, clip]{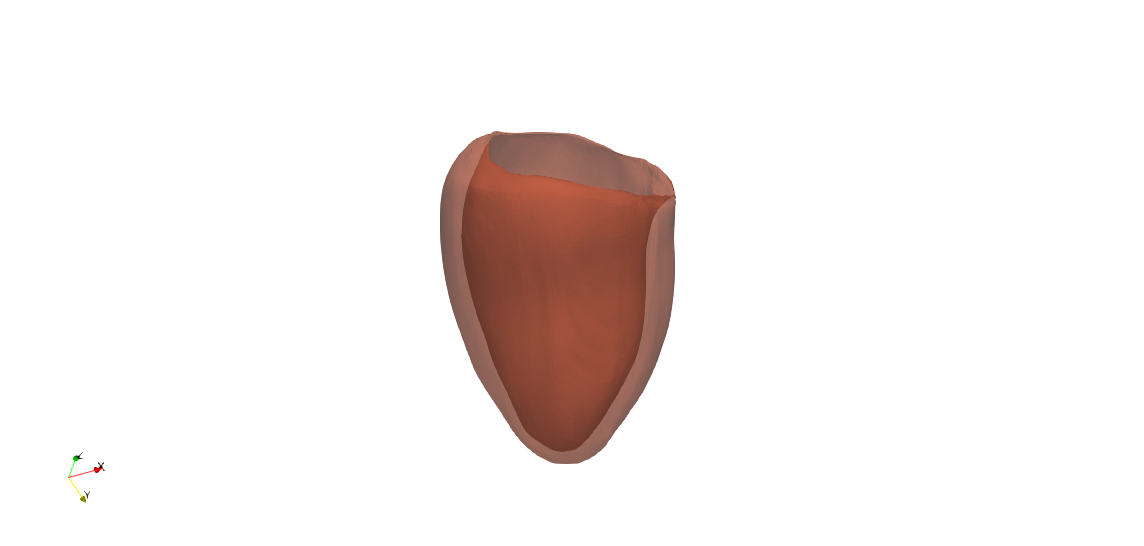} \\
  ~~~ &
  \raisebox{0.1\height}{\rotatebox[origin=c]{0}{\makecell{~\scalebox{0.8}{\textbf{t=0}}}}} &
  \raisebox{0.1\height}{\rotatebox[origin=c]{0}{\makecell{~\scalebox{0.8}{\textbf{t=10}}}}} &
  \raisebox{0.1\height}{\rotatebox[origin=c]{0}{\makecell{~\scalebox{0.8}{\textbf{t=20}}}}} &
  \raisebox{0.1\height}{\rotatebox[origin=c]{0}{\makecell{~\scalebox{0.8}{\textbf{t=30}}}}} &
  \raisebox{0.1\height}{\rotatebox[origin=c]{0}{\makecell{~\scalebox{0.8}{\textbf{t=40}}}}}
  \end{tabular}
  \caption{Motion tracking results across the cardiac cycle using the baseline methods and the proposed method.}
  \label{cycle_results}
\end{figure}

We compare the proposed method with baseline methods for the performance of motion estimation across the cardiac cycle. Fig.~\ref{cycle_results} demonstrates that MulViMotion~\cite{Meng2022_mulvimotion}, MeshMotion~\cite{Meng2022_miccai} and the proposed method are able to estimate both in-plane and through-plane motion while other methods only show motion within SAX plane. This is because~\cite{Meng2022_mulvimotion,Meng2022_miccai} and our method take full advantage of both SAX and LAX view images. Different from MulViMotion~\cite{Meng2022_mulvimotion} which estimates a voxel-wise motion field and generates 3D meshes from segmentations, the proposed method directly estimates the motion of each vertex on the heart mesh, and thus is able to keep the number of vertex and the vertex correspondences across the cardiac cycle. 
In contrast to MeshMotion~\cite{Meng2022_miccai} where the ED frame mesh of an individual heart is needed before motion tracking, the proposed method directly reconstructs the ED frame mesh by propagating from a template mesh. It integrates mesh reconstruction and mesh tracking into a single framework and also ensures the consistency of the meshes across different subjects. 
In addition, compared to~\cite{Meng2022_miccai}, we add a regularization loss $\mathcal{L}^{0\rightarrow t}_{reg}$ in this work to penalize the smoothness of the intermediate dense motion field ($\Phi_{0\rightarrow t}$). The results show that the proposed method achieves smoother LV basal part than~\cite{Meng2022_miccai}, \emph{e.g.}, in $t=20$ and $t=40$ frame in Fig.~\ref{cycle_results}.

\begin{figure*}[ptb]
 \centering
 \begin{tabular}{c@{\hspace{3\tabcolsep}}c@{\hspace{3\tabcolsep}}c@{\hspace{1.5\tabcolsep}}c@{\hspace{1\tabcolsep}}c@{\hspace{1\tabcolsep}}c@{\hspace{1\tabcolsep}}c}
  \includegraphics[height=2cm, trim=14cm 2cm 14cm 3cm, clip]{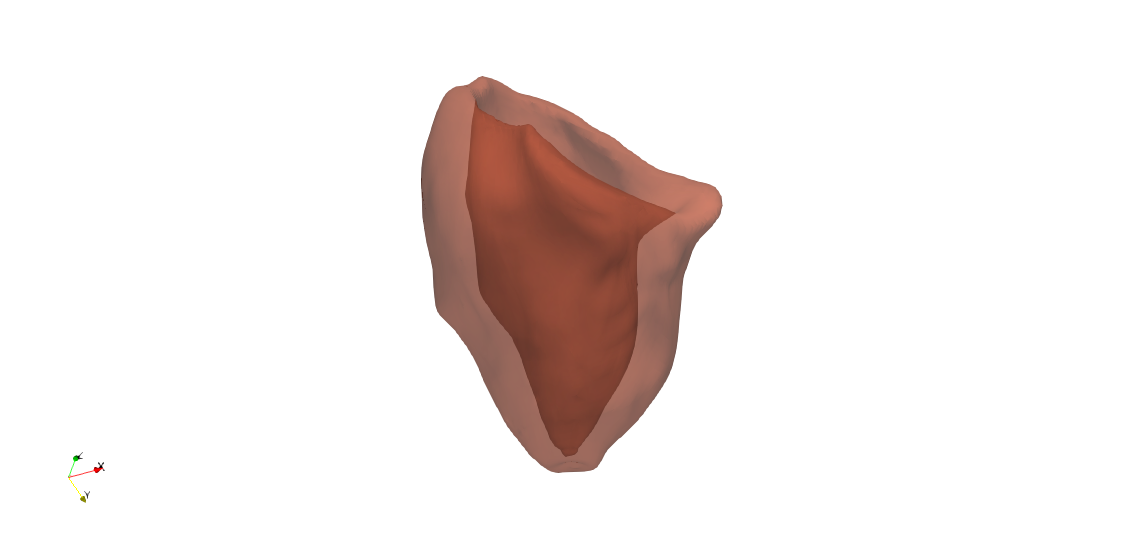} &
  \includegraphics[height=2cm, trim=14cm 2cm 14cm 3cm, clip]{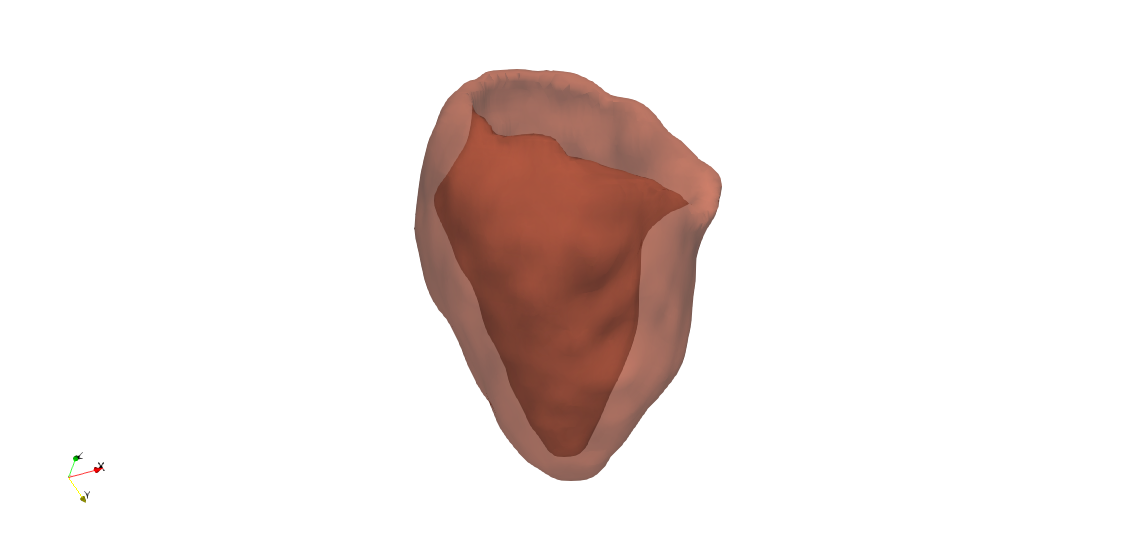} &
  \includegraphics[height=2cm, trim=14cm 2cm 14cm 3cm, clip]{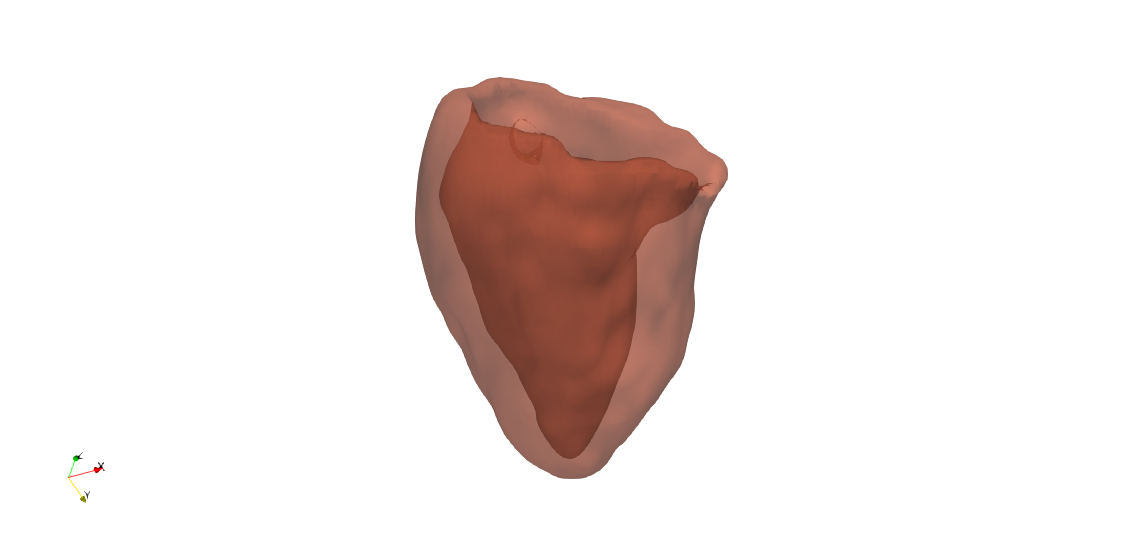} &
  \includegraphics[height=2cm, trim=14cm 2cm 14cm 3cm, clip]{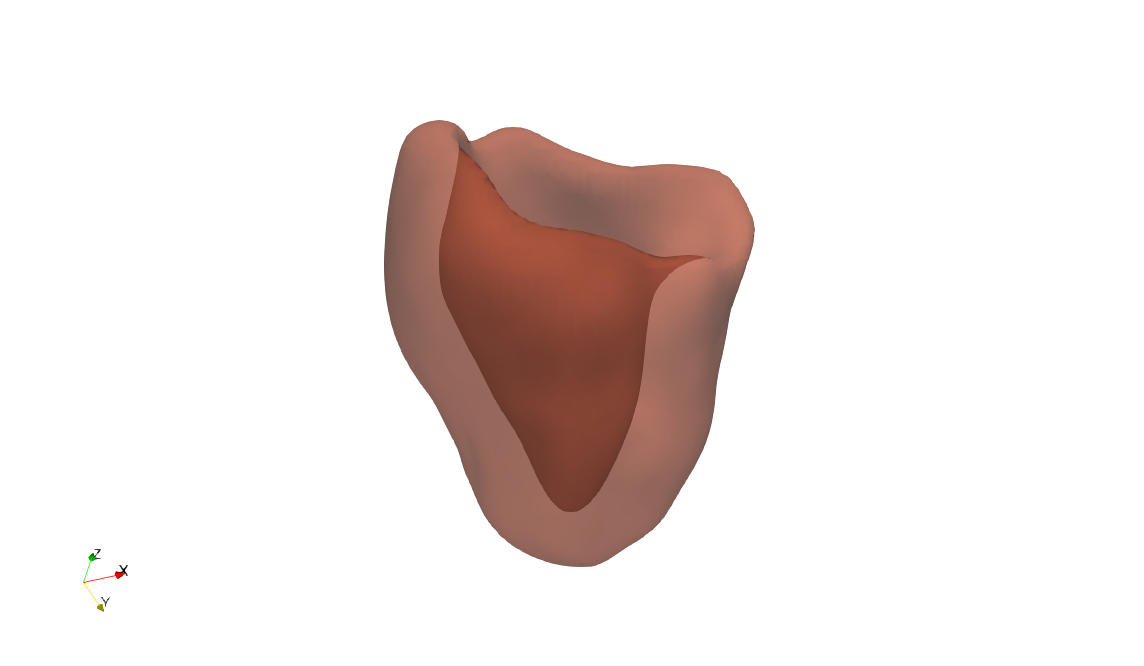} &
  \includegraphics[height=2cm, trim=14cm 2cm 14cm 3cm, clip]{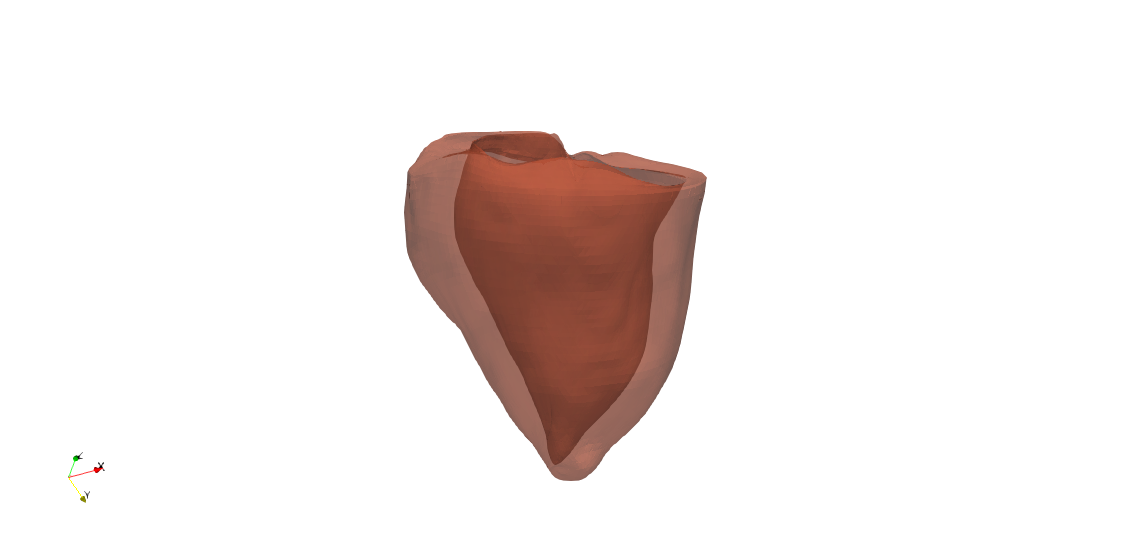} &
  \includegraphics[height=2cm, trim=14cm 2cm 14cm 3cm, clip]{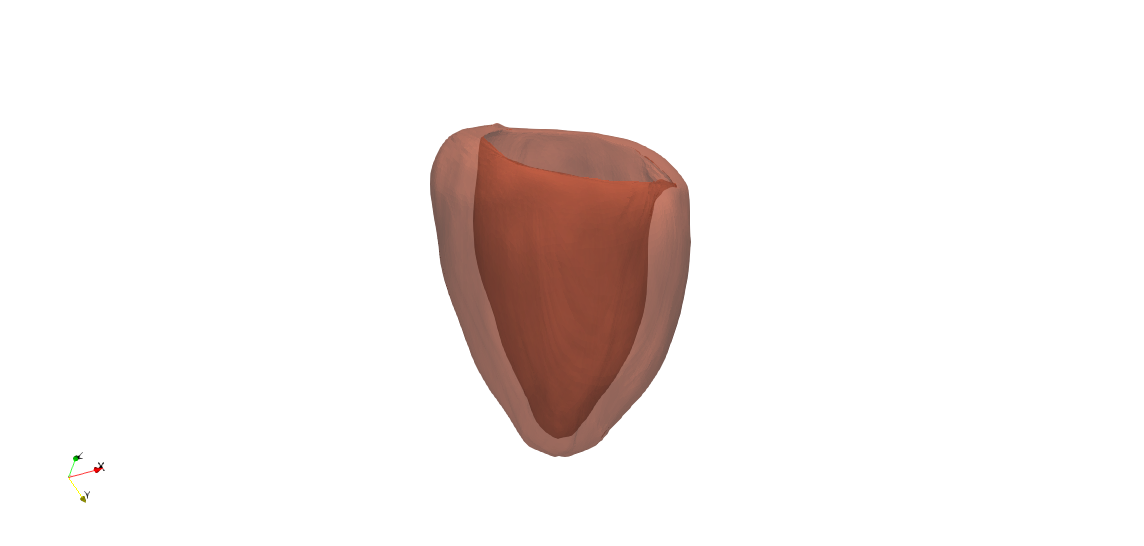} &
  \includegraphics[height=2cm, trim=14cm 2cm 14cm 3cm, clip]{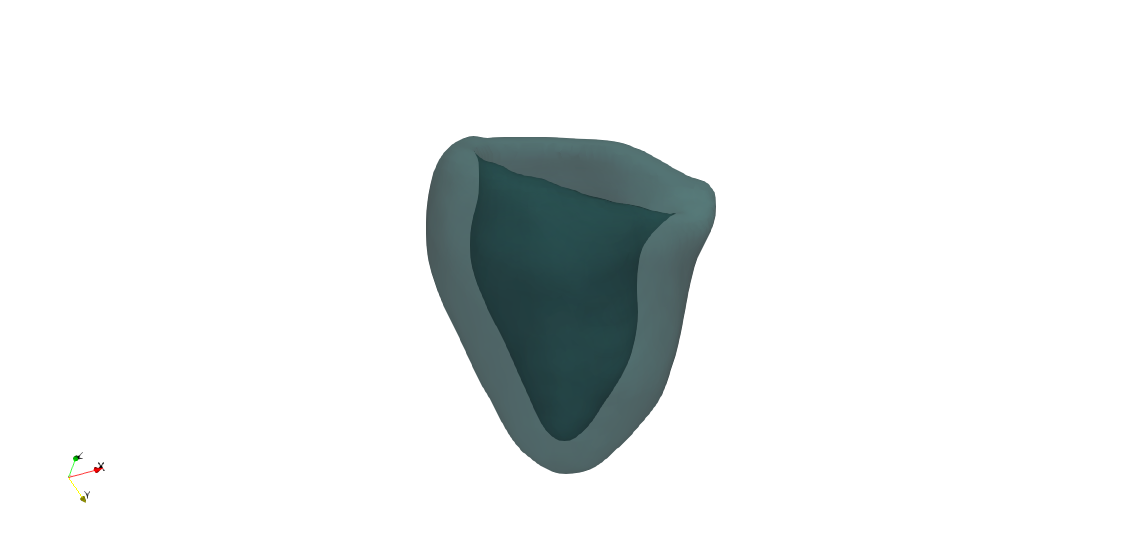} \\
  
  \raisebox{0.1\height}{\rotatebox[origin=c]{0}{\makecell{~\scalebox{0.8}{\textbf{FFD~\cite{Rueckert1999}}}}}} &
  \raisebox{0.1\height}{\rotatebox[origin=c]{0}{\makecell{~\scalebox{0.8}{\textbf{dDemons~\cite{Vercauteren2007}}}}}} &
  \raisebox{0.1\height}{\rotatebox[origin=c]{0}{\makecell{~\scalebox{0.8}{\textbf{3D-UNet~\cite{ociek2016}}}}}} &
  \raisebox{0.1\height}{\rotatebox[origin=c]{0}{\makecell{~\scalebox{0.8}{\textbf{MulViMotion~\cite{Meng2022_mulvimotion}}}}}} &
  \raisebox{0.1\height}{\rotatebox[origin=c]{0}{\makecell{~\scalebox{0.8}{\textbf{MeshMotion~\cite{Meng2022_miccai}}}}}} &
  \raisebox{0.1\height}{\rotatebox[origin=c]{0}{\makecell{~\scalebox{0.8}{\textbf{DeepMesh}}}}} &
  \raisebox{0.1\height}{\rotatebox[origin=c]{0}{\makecell{~\scalebox{0.8}{\textbf{ES frame (GT)}}}}}
  \end{tabular}
  \caption{Motion estimation using baseline methods and the proposed method. Green mesh is ground truth (GT) mesh of the ES frames. Red meshes are the predicted ES frame meshes based on different methods.}
  \label{mesh_comparison}
\end{figure*}

We further compare different methods by estimating the 3D motion field from ED frame to ES frame, which shows the largest deformation. Table~\ref{quantitative_comparison} shows the quantitative comparison results and Fig.~\ref{mesh_comparison} shows the qualitative results. From Table~\ref{quantitative_comparison}, we observe that the proposed method outperforms all baseline methods and achieves the best performance regarding SAX, 2CH and 4CH view segmentations. In addition, the proposed method obtains the ES frame mesh which is most similar to the ground truth ES frame mesh in Fig.~\ref{mesh_comparison}. These results demonstrate the effectiveness of the proposed method for estimating 3D mesh motion fields.

\begin{table*}[tb]
\centering
\caption{Comparison of other cardiac motion tracking methods. The results are reported as ``mean (standard deviation)". $\uparrow$ indicates the higher value the better while $\downarrow$ indicates the lower value the better. Best results in bold.}
\label{quantitative_comparison}
\resizebox{\textwidth}{!}{
\begin{tabular}{lcc|ccc|ccc}
\toprule[1.2pt]
\multirow{2}{*}{Methods}                        & 
\multirow{2}{*}{\makecell{Anatomical\\ views}}               &
\multirow{2}{*}{\makecell{Mean Surface \\ distance} $\downarrow$}               &
\multicolumn{3}{c|}{HD (mm) $\downarrow$ }            &
\multicolumn{3}{c}{BoundF ($\%$) $\uparrow$ }            \\
\cmidrule{4-9}
~~~~~ &
~~~~~ &
~~~~~ &
SAX &
2CH &
4CH &
SAX &
2CH &
4CH \\
\midrule
FFD~\cite{Rueckert1999}                        &
SAX                                              &
3.02(0.86) &
10.31 (3.55) &
15.17(4.52) &
15.95(4.84) &
62.15(7.48) &
77.60(6.97) &
77.79(7.13) \\
dDemons~\cite{Vercauteren2007}                            &
SAX                                  &
3.20(0.90) &
9.71 (4.07) &
15.01(3.48) &
15.72(3.41) &
63.67(6.92) &
77.38(5.99) &
80.29(5.83) \\
3D-UNet~\cite{ociek2016}                                &
SAX                                             &
3.35(0.88) &
8.88(3.88) &
14.44(2.99) &
14.83(3.57) &
60.64(7.74) &
74.63(6.01) &
76.06(6.08) \\
MulViMotion~\cite{Meng2022_mulvimotion}                               &
SAX, 2CH, 4CH                                              &
2.39(0.79) &
9.86(3.21) &
9.66(3.09) &
10.18(3.58) &
63.65(8.42) &
77.59(5.17) &
76.86(6.15) \\
MeshMotion~\cite{Meng2022_miccai}                                           &
SAX, 2CH, 4CH                                              &
1.98(0.44) &
9.73 (3.96) &
7.44(4.04) &
8.62(4.49) &
71.49(8.82) &
87.21(6.97)&
84.24(6.84) \\
DeepMesh                                           &
SAX, 2CH, 4CH                                              &
\textbf{1.66(0.51)} &
\textbf{9.08(3.86)} &
\textbf{5.75(1.81)} &
\textbf{6.21(2.56)} &
\textbf{74.95(8.25)} &
\textbf{89.26(6.97)} &
\textbf{88.69(6.23)} \\
\bottomrule[1.2pt]
\end{tabular}
}
\end{table*}

\subsubsection{Ablation study}
For the proposed method, we explore the effects of using different anatomical views and loss combinations in the mesh reconstruction and the mesh motion estimation. We utilize Hausdorff Distance (HD) for the evaluation. 
Table~\ref{reconstruction_diff} and Table~\ref{motionest_diff} show that adding the LAX view images improves the performance. This might be because LAX views can introduce high-resolution through-plane information for 3D motion estimation. These tables also show that proposed method with all the losses performs best in both mesh reconstruction and motion tracking, which illustrates the importance of each loss component.

\begin{table}[t]
  \begin{minipage}[t]{0.49\textwidth} 
    \caption{Mesh \textbf{reconstruction} with different anatomical views and different loss combinations.}
    \label{reconstruction_diff}
    \resizebox{\textwidth}{!}{
    \begin{tabular}{ccc|ccc}
    \toprule[1.2pt]
    \multicolumn{3}{c|}{Anatomical views}                 &
    \multicolumn{3}{c}{HD (mm) $\downarrow$ }            \\
    \midrule
    SAX &
    2CH &
    4CH &
    SAX &
    2CH &
    4CH \\
    \midrule
    $\surd$     &
    ~~~~~     &
    ~~~~~     &
    7.13 (2.25)&
    14.17(3.67) &
    14.36(3.54) \\
    $\surd$     &
    $\surd$    &
    ~~~~~     &
    6.11 (2.02) &
    7.06(2.61) &
    7.41(2.75) \\
    $\surd$     &
    ~~~~~    &
    $\surd$    &
    \textbf{5.50 (2.29)} &
    6.10(2.26) &
    6.14(2.32) \\
    $\surd$     &
    $\surd$    &
    $\surd$    &
    5.96 (2.14) &
    \textbf{5.97(1.83)} &
    \textbf{6.06(1.94)} \\
    \bottomrule[1.2pt]
    \end{tabular}
    }
  \end{minipage}%
  \vspace{0.1cm}
  \begin{minipage}[t]{0.49\textwidth}
    \resizebox{\textwidth}{!}{
    \begin{tabular}{cccc|ccc}
    \toprule[1.2pt]
    \multicolumn{4}{c|}{Loss combinations}                 &
    \multicolumn{3}{c}{HD (mm) $\downarrow$ }            \\
    \midrule
    $\mathcal{L}^{tpl\rightarrow 0}_{shape}$ &
    $\mathcal{L}^{tpl\rightarrow 0}_{smooth}$ &
    $\mathcal{L}_{surf}$ &
    $\mathcal{L}^{tpl\rightarrow 0}_{reg}$ &
    SAX &
    2CH &
    4CH \\
    \midrule
    $\surd$     &
    ~~~~~     &
    ~~~~~     &
    ~~~~~     &
    7.22 (2.15) &
    9.63(3.38) &
    9.52(3.02) \\
    $\surd$     &
    $\surd$    &
    ~~~~~     &
    ~~~~~     &
    10.95 (2.78) &
    14.26(3.34) &
    14.19(3.41) \\
    $\surd$     &
    $\surd$    &
    $\surd$    &
    ~~~~~    &
    6.58 (2.22) &
    10.62(5.18) &
    12.66(5.79) \\
    $\surd$     &
    $\surd$    &
    $\surd$    &
    $\surd$    &
    \textbf{5.96 (2.14)} &
    \textbf{5.97(1.83)} &
    \textbf{6.06(1.94)} \\
    \bottomrule[1.2pt]
    \end{tabular}
    }
  \end{minipage}
\end{table}

\begin{table}[t]
  \begin{minipage}[t]{0.49\textwidth} 
    \caption{Mesh \textbf{motion estimation} with different anatomical views and different loss combinations.}
    \label{motionest_diff}
    \resizebox{\textwidth}{!}{
    \begin{tabular}{ccc|ccc}
    \toprule[1.2pt]
    \multicolumn{3}{c|}{Anatomical views}                 &
    \multicolumn{3}{c}{HD (mm) $\downarrow$ }            \\
    \midrule
    SAX &
    2CH &
    4CH &
    SAX &
    2CH &
    4CH \\
    \midrule
    $\surd$     &
    ~~~~~     &
    ~~~~~     &
    \textbf{8.49 (3.67)
} &
    6.11(2.09) \  &
    7.31(2.39) \  \\
    $\surd$     &
    $\surd$    &
    ~~~~~     &
    8.50 (3.54) &
    5.88(2.41) &
    6.79(3.22) \\
    $\surd$     &
    ~~~~~    &
    $\surd$    &
    9.38 (4.17) &
    6.37(1.78) &
    6.27(2.29) \\
    $\surd$     &
    $\surd$    &
    $\surd$    &
    9.08(3.86) &
    \textbf{5.75(1.81)} &
    \textbf{6.21(2.56)} \\
    \bottomrule[1.2pt]
    \end{tabular}
    }
  \end{minipage}%
  \vspace{0.1cm}
  \begin{minipage}[t]{0.49\textwidth}
    \resizebox{\textwidth}{!}{
    \begin{tabular}{cccc|ccc}
    \toprule[1.2pt]
    \multicolumn{4}{c|}{Loss combinations}                 &
    \multicolumn{3}{c}{HD (mm) $\downarrow$ }            \\
    \midrule
    $\mathcal{L}^{0\rightarrow t}_{shape}$ &
    $\mathcal{L}^{0\rightarrow t}_{smooth}$ &
    $\mathcal{L}_{sim}$ &
    $\mathcal{L}^{0\rightarrow t}_{reg}$ &
    SAX &
    2CH &
    4CH \\
    \midrule
    $\surd$     &
    ~~~~~     &
    ~~~~~     &
    ~~~~~     &
    \textbf{8.97 (3.35)
} &
    7.20(2.00) \  &
    7.34(2.13) \  \\
    $\surd$     &
    $\surd$    &
    ~~~~~     &
    ~~~~~     &
    11.08 (4.07) &
    10.56(2.23) \  &
    10.43(2.26) \  \\
    $\surd$     &
    $\surd$    &
    $\surd$    &
    ~~~~~    &
    9.92 (3.44) &
    8.19(2.52) \  &
    8.67(3.21) \  \\
    $\surd$     &
    $\surd$    &
    $\surd$    &
    $\surd$    &
    9.08(3.86) &
    \textbf{5.75(1.81)} &
    \textbf{6.21(2.56)} \\
    \bottomrule[1.2pt]
    \end{tabular}
    }
  \end{minipage}
\end{table}

\subsubsection{The influence of hyper-parameters}
\begin{figure}[t]
    \centering
    \setcounter{subfigure}{0}
    \hspace{-0.5cm}
    \subfloat[Various $\lambda_1$, $\beta_1$, $\gamma_1$]{
    \begin{tabular}{c}
         \includegraphics[height=3.5cm, trim=0cm 0cm 1cm 0.5cm, clip]{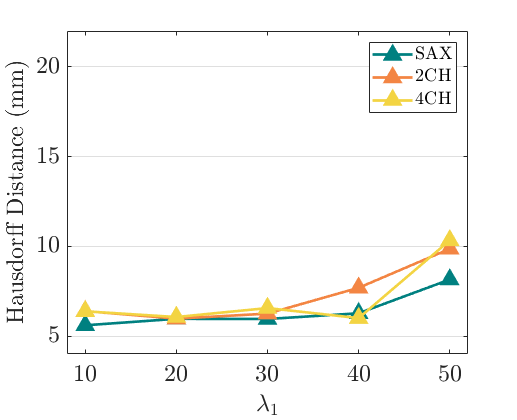} \\
         \includegraphics[height=3.5cm, trim=0cm 0cm 1cm 0.5cm, clip]{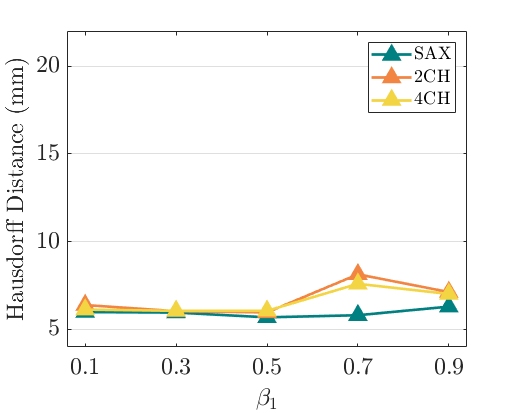} \\
         \includegraphics[height=3.5cm, trim=0cm 0cm 1cm 0.5cm, clip]{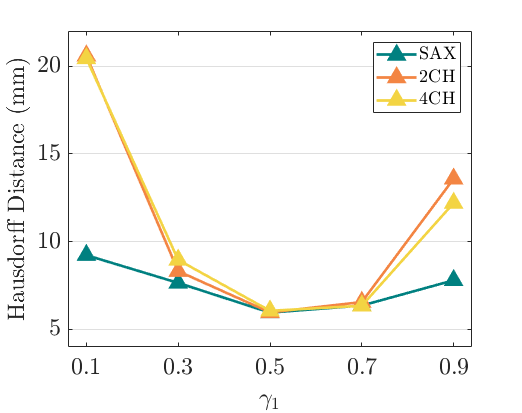}
    \end{tabular}
    }
    \hspace{-0.9cm}
    \setcounter{subfigure}{1}
    \subfloat[Various $\lambda_2$, $\beta_2$, $\gamma_2$]{
    \begin{tabular}{c}
         \includegraphics[height=3.5cm, trim=0cm 0cm 1cm 0.5cm, clip]{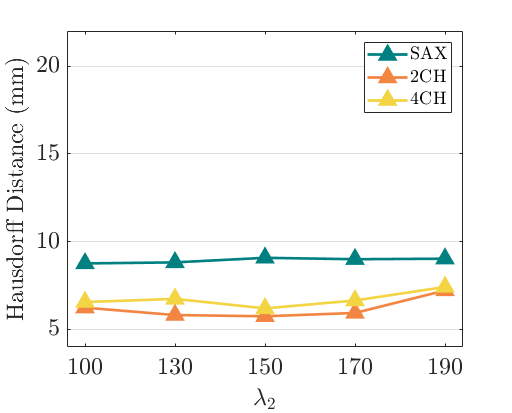}  \\
         \includegraphics[height=3.5cm, trim=0cm 0cm 1cm 0.5cm, clip]{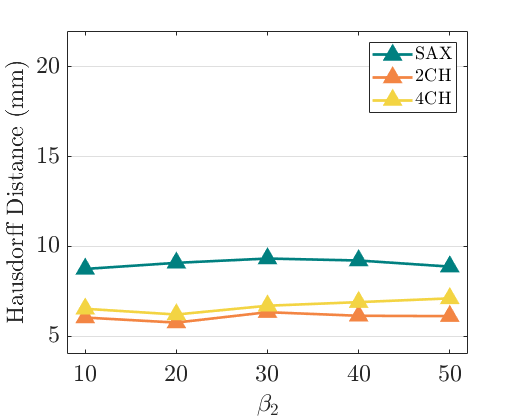}  \\
         \includegraphics[height=3.5cm, trim=0cm 0cm 1cm 0.5cm, clip]{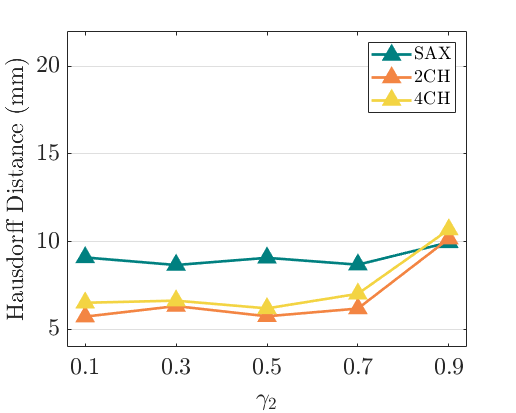}
    \end{tabular}
    } 
    \caption{Effects of varying hyper-parameters on Hausdorff distance. (a) shows the results of using various $\lambda_1$, $\beta_1$, $\gamma_1$ for the mesh reconstruction. The final selection is $\lambda_1=20, \beta_1=0.5, \gamma_1=0.5$. (b) shows the results of using various $\lambda_2$, $\beta_2$, $\gamma_2$ for the mesh motion tracking. The final selection is $\lambda_2=150, \beta_2=20, \gamma_2=0.5$. Note that when one hyper-parameter changes, other hyper-parameters are fixed to the final selection value.}
    \label{hyperpara}
\end{figure}

We evaluate the performance of mesh reconstruction and mesh motion estimation under various values of the hyper-parameters. Specifically, we compute Hausdorff distance (HD) based on the predicted and ground truth 2D myocardium contours on SAX, 2CH and 4CH view planes. We compare the contours of the ED frame for the mesh reconstruction and compare the contours of the ES frame for the mesh motion estimation.
Fig.~\ref{hyperpara} shows that in contrast to LAX views, the performance on the SAX view is not sensitive to hyper-parameters. This might because the SAX stacks contain multiple slices while the 2CH and 4CH view only have a single slice for evaluation. From the last row in Fig.~\ref{hyperpara} (a), we observe that a weak or a strong regularization on voxel-wise displacement may reduce the accuracy of mesh reconstruction.

\section{Discussion}


In the mesh motion estimation framework presented in this work, we predict the motion field of the heart mesh by sampling from an intermediate voxel-wise 3D motion field. An alternative to our method would be to estimate mesh motion fields directly from input images via fully connected layers without intermediate voxel-wise 3D motion estimation. However, using fully connected layers to estimate the displacement of $\sim 20K$ vertices needs large GPU memory, which may not always be available.



We use the weighted Hausdorff Distance to compare the extracted 2D contours and the ground truth 2D contours of myocardium in $\mathcal{L}^{tpl\rightarrow 0}_{shape}$ and $\mathcal{L}^{0\rightarrow t}_{shape}$. Other boundary similarity measurements that can evaluate the distance between soft-labeled and hard-labeled point sets may also be applied to this loss component in our task.

When evaluating motion estimation, we quantitatively evaluated the performance on the ES frame. This is because 3D ground truth meshes are only available at the ED and ES frames in our current dataset. More importantly, ES frame has the largest deformation from the ED frame, which is the most challenging case in motion estimation. Besides, using the ES frame for quantitative evaluation is same as other previous works, such as~\cite{Qin2018, Qin2020, YuH2020}.


We separately train the mesh reconstruction module and the mesh motion estimation module during training but the proposed method is end-to-end trainable. The probability
map (2D soft contours) obtained from the differentiable mesh-to-image rasterizer enables the differentiability of the rasterization. However, simultaneously training mesh reconstruction and mesh motion estimation may increase the complexity of hyper-parameter tuning.

\begin{figure}[t]
    \centering
    \setcounter{subfigure}{0}
    \hspace{-1cm}
    \subfloat[Before pre-processing]{
    \begin{tabular}{c}
         \includegraphics[height=2.5cm, trim=15cm 3cm 7cm 1.5cm, clip]{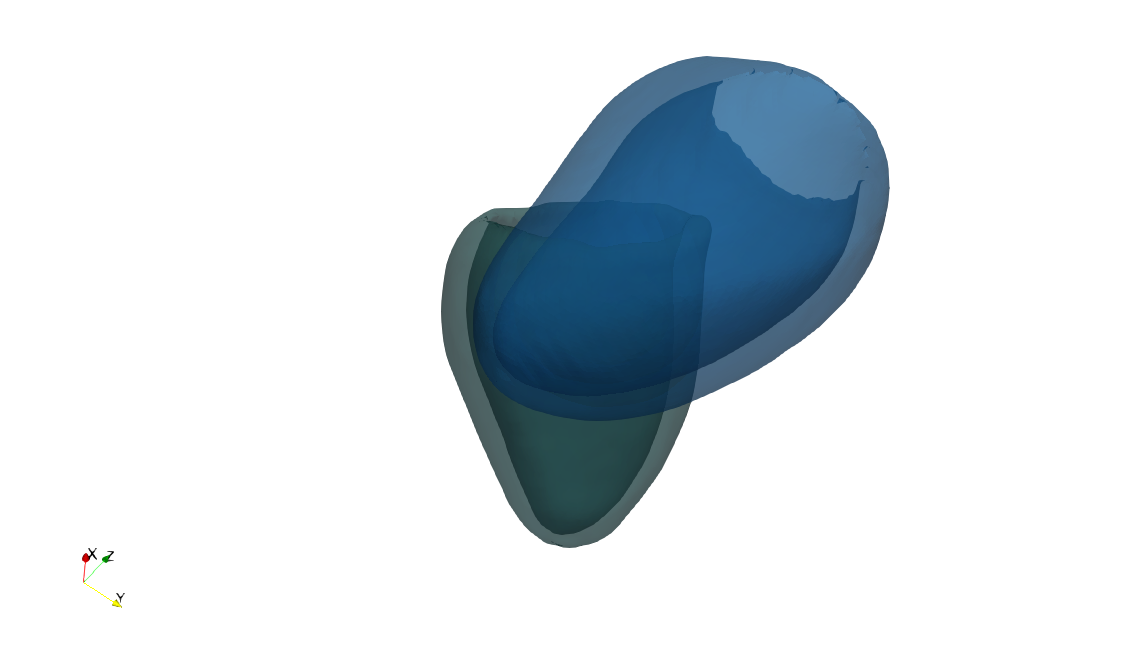} 
    \end{tabular}
    }
    \hspace{0.2cm}
    \setcounter{subfigure}{1}
    \subfloat[After pre-processing]{
    \begin{tabular}{c}
         \includegraphics[height=2.5cm, trim=15cm 3cm 7cm 1.5cm, clip]{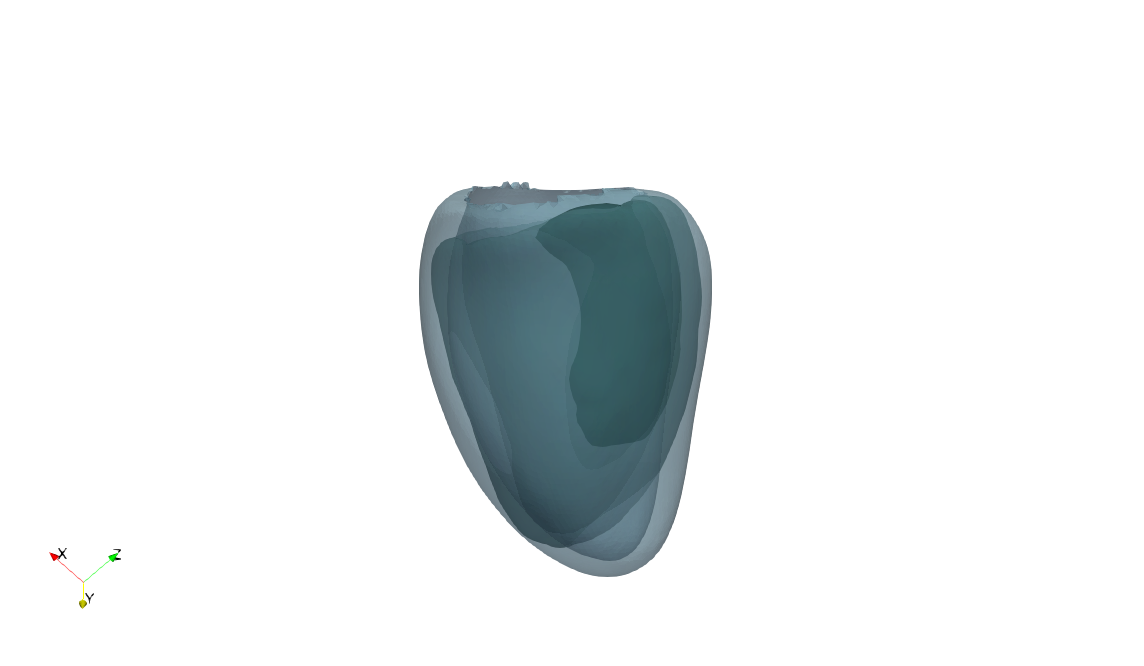} 
    \end{tabular}
    } 
    \hspace{-0.8cm}
    \caption{Comparison of the template and a subject ED frame mesh. (a) shows that the template is not in the same space as the subject mesh. (b) demonstrates that we can move the template to the subject space after data pre-processing. Green meshes are the ground truth subject mesh. Blue meshes are the template before and after data pre-processing.}
    \label{tplsub}
\end{figure}

The proposed deep neural network in the mesh reconstruction module focuses on deforming the template mesh to the ED frame mesh of individual subjects. To move the template mesh to individual subject space before mesh reconstruction, we utilize the information about the relative position in the DICOM header of 2D images. Fig.~\ref{tplsub} shows an example of moving the template mesh to a subject space during data pre-processing.

Our evaluation has been conducted on LV myocardial motion tracking because it is important for clinical assessment of cardiac function. However, the proposed method is not limited to LV myocardium. Our model can be easily adapted to 3D right ventricular myocardial motion tracking by using the corresponding template mesh and the ground truth 2D contours during training.


\begin{figure}[tb]
 \centering
 \begin{tabular}{@{\hspace{-1\tabcolsep}}c@{\hspace{0.1\tabcolsep}}c@{\hspace{0.1\tabcolsep}}c@{\hspace{0.1\tabcolsep}}c@{\hspace{0.1\tabcolsep}}c}
  \raisebox{0.5\height}{\rotatebox[origin=c]{90}{\makecell{~\scalebox{0.8}{\textbf{\makecell{Mesh \\ reconstruction}}}}}} &
  \includegraphics[height=1.7cm, trim=14cm 2cm 14cm 3.5cm, clip]{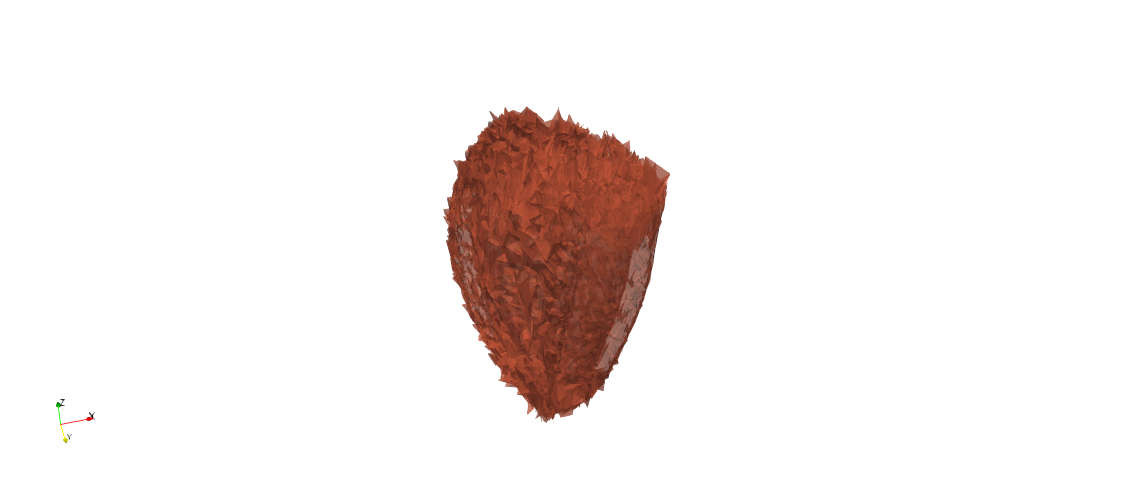} &
  \includegraphics[height=1.7cm, trim=14cm 2cm 14cm 3.5cm, clip]{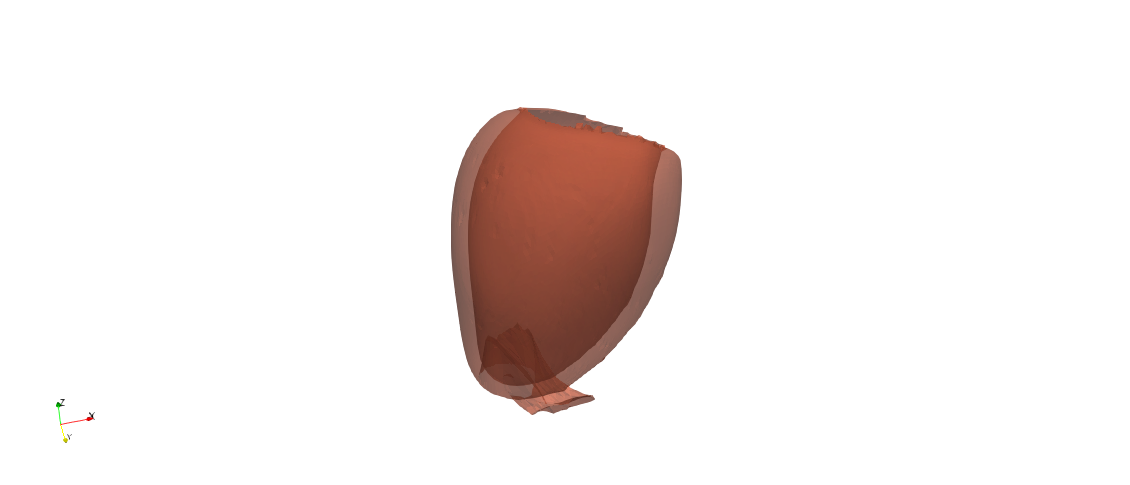} &
  \includegraphics[height=1.7cm, trim=14cm 2cm 14cm 3.5cm, clip]{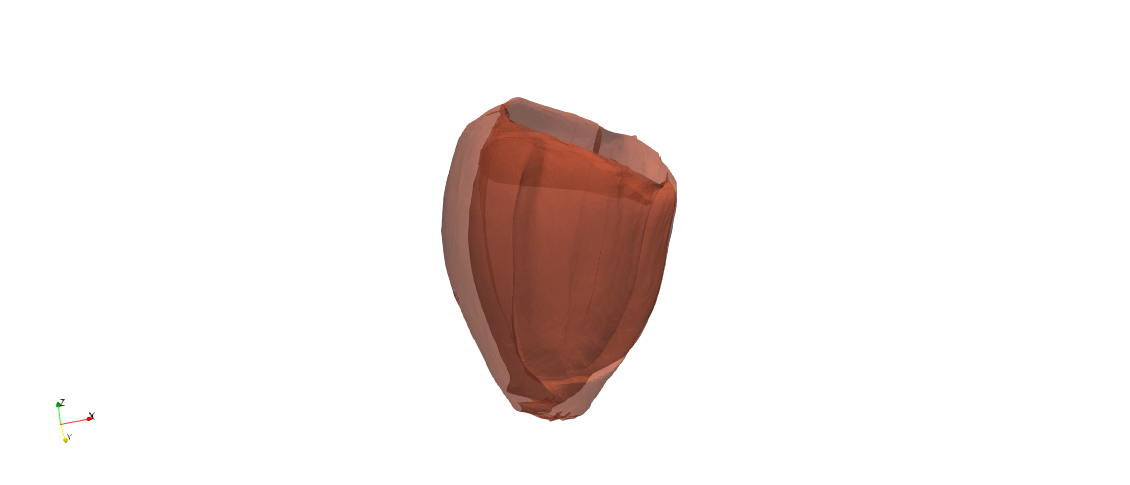} &
  \includegraphics[height=1.7cm, trim=14cm 2cm 14cm 3.5cm, clip]{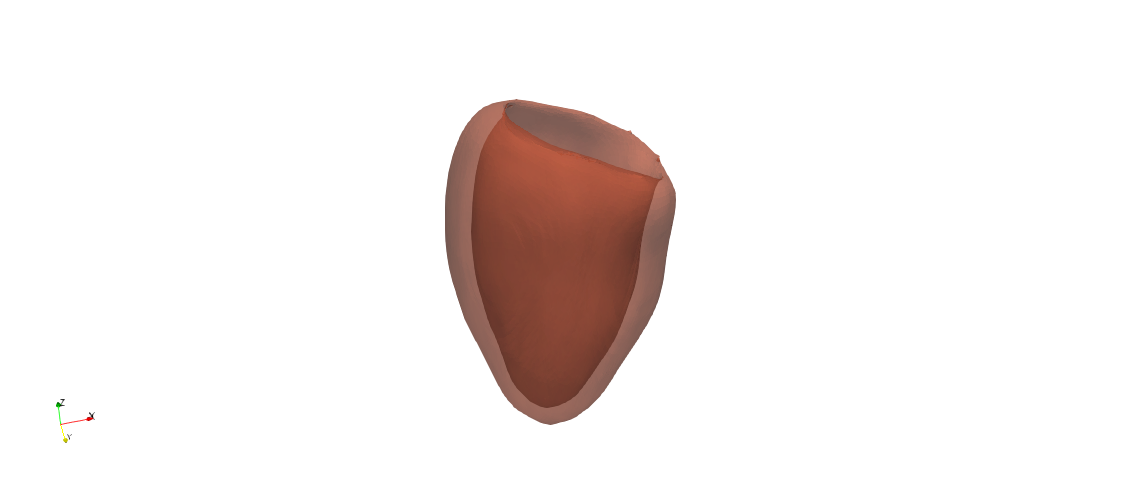} \\
  ~~~ &
  \raisebox{0.1\height}{\rotatebox[origin=c]{0}{\makecell{~\scalebox{0.6}{\textbf{$\mathcal{L}^{tpl\rightarrow 0}_{shape}$}}}}} &
  \raisebox{0.1\height}{\rotatebox[origin=c]{0}{\makecell{~\scalebox{0.6}{\textbf{$\mathcal{L}^{tpl\rightarrow 0}_{shape}+\mathcal{L}^{tpl\rightarrow 0}_{smooth}$}}}}} &
  \raisebox{0.1\height}{\rotatebox[origin=c]{0}{\makecell{~\scalebox{0.6}{\textbf{$\mathcal{L}^{tpl\rightarrow 0}_{shape}+\mathcal{L}^{tpl\rightarrow 0}_{smooth}+\mathcal{L}_{surf}$}}}}} &
  \raisebox{0.1\height}{\rotatebox[origin=c]{0}{\makecell{~\scalebox{0.6}{All losses}}}} \\
  \raisebox{0.6\height}{\rotatebox[origin=c]{90}{\makecell{~\scalebox{0.8}{\textbf{\makecell{Motion \\ estimation}}}}}} &
  \includegraphics[height=1.7cm, trim=14cm 2cm 14cm 3.5cm, clip]{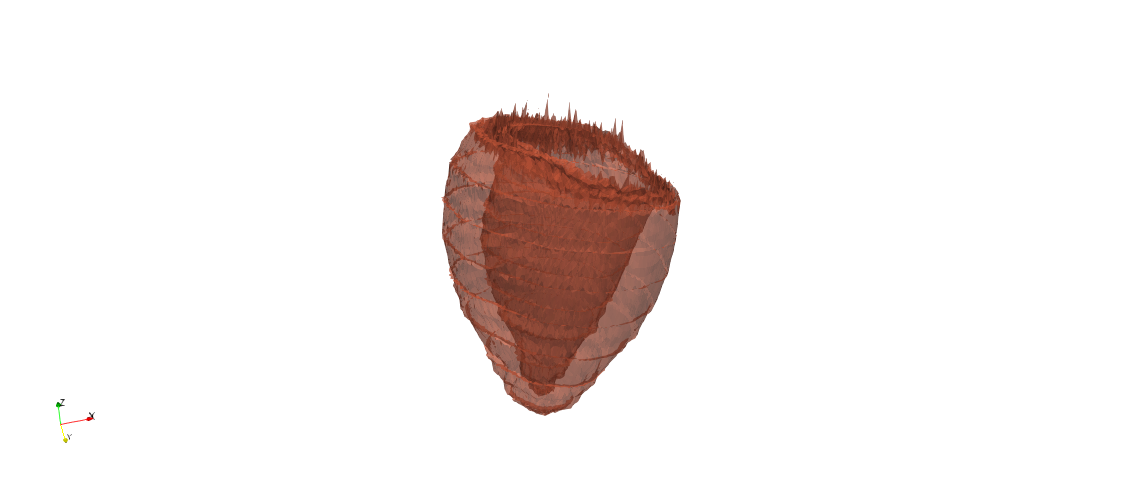} &
  \includegraphics[height=1.7cm, trim=14cm 2cm 14cm 3.5cm, clip]{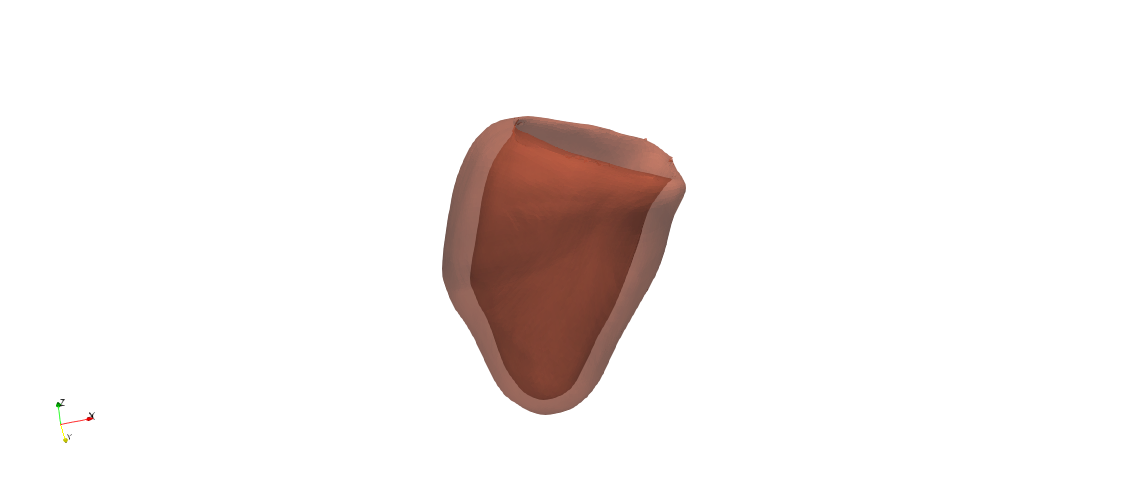} &
  \includegraphics[height=1.7cm, trim=14cm 2cm 14cm 3.5cm, clip]{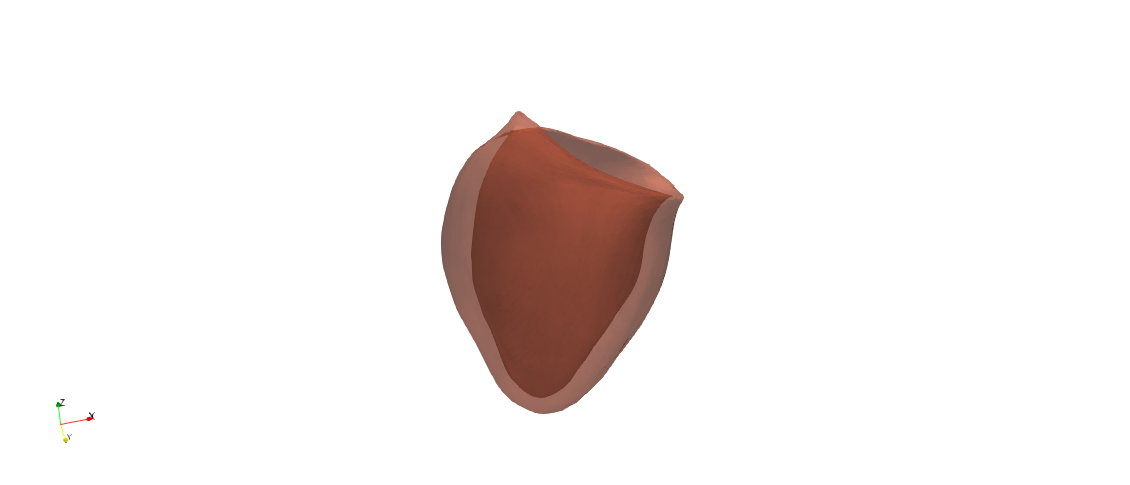} &
  \includegraphics[height=1.7cm, trim=14cm 2cm 14cm 3.5cm, clip]{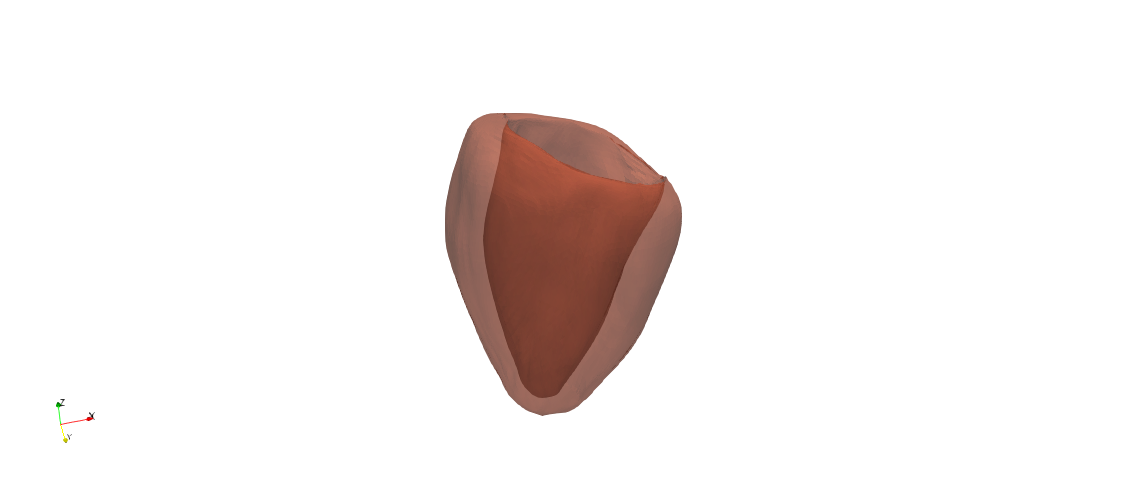} \\
  ~~~ &
  \raisebox{0.1\height}{\rotatebox[origin=c]{0}{\makecell{~\scalebox{0.6}{\textbf{$\mathcal{L}^{0\rightarrow t}_{shape}$}}}}} &
  \raisebox{0.1\height}{\rotatebox[origin=c]{0}{\makecell{~\scalebox{0.6}{\textbf{$\mathcal{L}^{0\rightarrow t}_{shape}+\mathcal{L}^{0\rightarrow t}_{smooth}$}}}}} &
  \raisebox{0.1\height}{\rotatebox[origin=c]{0}{\makecell{~\scalebox{0.6}{\textbf{$\mathcal{L}^{0\rightarrow t}_{shape}+\mathcal{L}^{0\rightarrow t}_{smooth}+\mathcal{L}_{sim}$}}}}} &
  \raisebox{0.1\height}{\rotatebox[origin=c]{0}{\makecell{~\scalebox{0.6}{All losses}}}}
  \end{tabular}
  \caption{Qualitative results of mesh reconstruction and mesh motion estimation with different combinations of losses. The top row shows the reconstructed ED frame mesh. The bottom row shows the estimated ES frame mesh.}
  \label{ablation_qualitative}
\end{figure}

Table III and Table IV show that only using shape regularization ($\mathcal{L}^{tpl\rightarrow 0}_{shape}$ and $\mathcal{L}^{0\rightarrow t}_{shape}$) achieves second best quantitative results. However, Fig.~\ref{ablation_qualitative} demonstrates that shape regularization alone is insufficient for good qualitative results while other regularization terms make contributions as well, to surface smoothness, accurate deformation and deformation smoothness, accordingly.

The proposed method is trained and evaluated on healthy subjects, where we aim to demonstrate the effectiveness of the methodology. We acknowledge that the current trained model may not achieve best performance on pathological data, especially hearts with specific diseases. To address this limitation, one possible solution is to include more pathological cases in the training set and re-train the model. In addition, there can be large deformation between the template mesh and pathological hearts, for which we may need to add extra regularization terms to the template-based mesh reconstruction module.

We believe that our mesh-based motion tracking method can benefit a variety of clinical applications. The proposed model can provide an accurate and holistic estimation of 3D geometry and motion, which could be used for clinical prediction and association tasks where conventional metrics provide weak discrimination. For example, we can model the association between cardiac motion (either globally, or vertex-wise) and demographics (\emph{e.g.}, age, gender), genetic predisposition, and disease risk factors. In particular, as our method maintains an anatomical correspondence of the cardiac meshes (\emph{i.e.}, the number of vertices and faces) in the cohort, it can facilitate learning complex motion features for specific tasks in a population. This could enable the use of motion-related traits for early diagnosis or  monitoring of disease progression. In addition, our method can support biophysical modeling by using meshes as input for mechanical simulations. This can potentially improve our understanding of cardiac physiology. Also, the predicted sequence of meshes can be used for computing conventional volumetric and functional biomarkers (\emph{e.g.}, ED volume, ejection fraction). However, in many existing clinical studies, volumetric biomarkers are computed from segmentations. To gain clinical acceptance, complex motion based traits derived from mesh-based methods will need to be thoroughly validated against the conventional segmentation-based metrics used in current  practice.

\section{Conclusion}
In this paper, we propose a novel deep learning method for template-guided mesh-based cardiac motion tracking. The proposed method reconstructs the 3D heart mesh of the reference frame and estimate per-vertex motion field from 2D SAX and LAX view CMR images. 
The proposed method enables both mesh reconstruction and mesh motion tracking.
It is also capable of maintaining the number of vertices and vertex correspondences across the cardiac cycle.
Experimental results demonstrate the effectiveness of the proposed method compared with other competing methods.

\bibliographystyle{abbrv}
\balance
\bibliography{reference}

\end{document}